\documentclass[journal]{IEEEtran}

\usepackage{amsmath,graphicx,amssymb}
\makeatletter
\def\maketag@@@#1{\hbox{\m@th\normalfont\normalsize#1}}
\makeatother

\usepackage{cite}
\usepackage{tikz}
\usepackage{mwe}
\usepackage{multicol}

\usepackage{caption}
\usepackage{subcaption}

\usepackage{lipsum}
\usepackage[normalem]{ulem}
\useunder{\uline}{\ul}{}
\usetikzlibrary{arrows}
\usepackage{multirow}
\usepackage[font=small,skip=1pt]{caption}
\usepackage{booktabs}

\usetikzlibrary{intersections}
\usepackage{tkz-euclide}
\usetkzobj{all}

\usepackage{pgfplots}
\usepackage{balance}

\usepackage{algorithmic}
\usepackage{algorithm}

\DeclareMathOperator*{\argmin}{arg\,min}
\DeclareMathOperator*{\argmax}{arg\,max}
\newcommand\myeq{\stackrel{\mathclap{\normalfont\mbox{c}}}{=}}

\usepackage{mathtools}
\DeclarePairedDelimiter\floor{\lfloor}{\rfloor}

\ifCLASSINFOpdf
\else
\fi

\begin{document}
%
\title{Model-based Speech Enhancement for Intelligibility Improvement in Binaural Hearing Aids }
%
%
%

\author{Mathew~Shaji~Kavalekalam,~\IEEEmembership{Student Member,~IEEE,} Jesper Kj\ae r Nielsen, \IEEEmembership{Member, IEEE,}
        Jesper B\"unsow Boldt,~\IEEEmembership{Member, IEEE}
        and~Mads Gr\ae sb\o ll Christensen,~\IEEEmembership{Senior Member, IEEE}
\thanks{Mathew S. Kavalekalam, Jesper K. Nielsen and Mads G. Christensen are with the Audio Analysis Lab, Department of Architecture, Design and Media Technology at Aalborg University.}
\thanks{Jesper Boldt is with GN Hearing, Ballerup, Denmark}
\thanks{Manuscript received ; revised }}

%
%

\markboth{Journal of \LaTeX\ Class Files,~Vol.~14, No.~XX, X~20XX}%
{Shell \MakeLowercase{\textit{et al.}}: Bare Demo of IEEEtran.cls for IEEE Journals}
%



\maketitle

\begin{abstract}
Speech intelligibility is often severely degraded among hearing impaired individuals in situations such as the cocktail party scenario. The performance of the current hearing aid technology has been observed to be limited in these scenarios. In this paper, we propose a binaural speech enhancement framework that takes into consideration the speech production model. The enhancement framework proposed here is based on the Kalman filter that allows us to take the speech production dynamics into account during the enhancement process. The usage of a Kalman filter requires the estimation of clean speech and noise short term predictor (STP) parameters, and the clean speech pitch parameters. In this work, a binaural codebook-based method is proposed for estimating the STP parameters, and a directional pitch estimator based on the harmonic model and maximum likelihood principle is used to estimate the pitch parameters. The proposed method for estimating the STP and pitch parameters jointly uses the information from  left and right ears, leading to a more robust estimation of the filter parameters. Objective measures such as PESQ and STOI have been used to evaluate the enhancement framework in different acoustic scenarios representative of the cocktail party scenario. We have also conducted subjective listening tests on a set of nine normal hearing subjects, to evaluate the performance in terms of intelligibility and quality improvement. The listening tests show that the proposed algorithm, even with access to only a single channel noisy observation, significantly improves the overall speech quality, and the speech intelligibility by up to $15\%$.
\end{abstract}


%
\IEEEpeerreviewmaketitle

\section{Introduction}

 Normal hearing (NH) individuals have the ability to concentrate
on a single speaker even in the presence of multiple interfering speakers. This phenomenon is termed as the cocktail party
effect. However, hearing impaired individuals lack this ability to separate out a single speaker in the presence of
multiple competing speakers. This leads to listener fatigue and isolation of the hearing aid (HA) user. Mimicking the cocktail party effect in a digital HA is very much desired in such scenarios \cite{kochkin200210}. Thus, to help the HA user to focus on a particular speaker,
speech enhancement has to be performed to reduce the effect of the interfering speakers. The primary objectives of a speech enhancement system in HA are to improve the intelligibility and quality of the degraded speech. Often, a hearing impaired person is fitted with HAs at both ears. Modern HAs have the technology to wirelessly communicate with each other making it possible to share information between the HAs. Such a property in HAs enables the use of binaural speech enhancement algorithms. The binaural processing of noisy signals has shown to be more effective than processing the noisy signal independently at each ear due to the utilization of spatial information \cite{van2009speech}. Apart from a better noise reduction performance, binaural algorithms make it possible to preserve the binaural cues which contribute to spatial release from masking \cite{bronkhorst1988effect}. 
Often, HAs are fitted with multiple microphones at both ears. Some binaural speech enhancement algorithms developed for such cases are \cite{doclo2008acoustic, cornelis2010theoretical}. In \cite{doclo2008acoustic}, a multichannel Wiener filter for HA applications is proposed which results in a minimum mean squared error (MMSE) estimation of the target speech. These methods were shown to distort the binaural cues of the interfering noise while maintaining the binaural cues of the target. Consequently, a method was proposed in \cite{klasen2007binaural} that introduced a parameter to trade off between the noise reduction and cue preservation. The above mentioned algorithms have reported improvements in speech intelligibility.  

We are here mainly concerned with the binaural enhancement of speech with access to only one microphone per HA \cite{dorbecker1996combination, li2011two, lotter2006dual}. More specifically, this paper is concerned with a two-input two-output system. This situation is encountered in in-the-ear (ITE) HAs, where the space constraints limit the number of microphones per HA. Moreover, in the case where we have multiple microphones per HA, beamforming can be applied individually on each HA to form the two inputs, which can then be processed further by the proposed dual channel enhancement framework. 
One of the first approaches to perform dual channel speech enhancement was that of \cite{dorbecker1996combination} where a two channel spectral subtraction was combined with an adaptive Wiener post-filter. This led to a distortion of the binaural cues, as different gains were applied to the left and right channels. Another approach to performing dual channel speech enhancement was proposed in \cite{li2011two} and this solution consisted of two stages. The first stage dealt with the estimation of interference signals using an equalisation-cancellation theory, and the second stage was an adaptive Wiener filter. The intelligibility improvements corresponding to the algorithms stated above have not been studied well. These algorithms perform the enhancement in the frequency domain by assuming that the speech and noise components are uncorrelated, and do not take into account the nature of the speech production process. 
In this paper, we propose a binaural speech enhancement framework that takes the speech production model into account. The model used here is based on the source-filter model, where the filter corresponds to the vocal tract and the source corresponds to the excitation signal produced by the vocal chords. Using a physically meaningful model gives us a sufficiently accurate way for explaining how the signals were generated, but also helps in reducing the number of parameters to be estimated. One way to exploit this speech production model for the enhancement process is to use a Kalman filter, as the speech production dynamics can be modelled within the Kalman filter using the state space equations while also accounting for the background noise. Kalman filtering for single channel speech enhancement in the presence of white background noise was first proposed in \cite{paliwal1987kalman}. This work was later extended to deal with coloured noise in \cite{gibson1991filtering,gannot1998iterative}. One of the main limitations of Kalman filtering based enhancement is that the state space parameters required for the formulation of the state space equations need to be known or estimated. The estimation of the state space parameters is a difficult problem due to the non-stationary nature of speech and the presence of noise. The state space parameters are the autoregressive (AR) coefficients and the excitation variances for the speech and noise respectively. Henceforth, AR coefficients along with the excitation variances will be denoted as the short term predictor (STP) parameters. In \cite{gibson1991filtering,gannot1998iterative} these STP parameters  were estimated using an approximated expectation-maximisation algorithm. However, the performance of these algorithms were noted to be unsatisfactory in non-stationary noise environments. Moreover, these algorithms assumed the excitation signal in the source-filter model to be white Gaussian noise. Even though this assumption is appropriate for modelling unvoiced speech, it is not very suitable for modelling voiced speech. This issue was handled in \cite{goh1999kalman} by using a modified model for the excitation signal capable of modelling both voiced and unvoiced speech. The usage of this model for the enhancement process required the estimation of the pitch parameters in addition to the STP parameters. This modification of the excitation signal was found to improve the performance in voiced speech regions, but the performance of the algorithm in the presence of non-stationary background noise was still observed to be unsatisfactory. This was primarily due to the poor estimation of the model parameters in non-stationary background noise. The noise STP parameters were estimated in \cite{goh1999kalman} by assuming that the first 100 milli seconds of the speech segment contained only noise and the parameters were then assumed to be constant.

In this work, we introduce a binaural model-based speech enhancement framework which addresses the poor estimation of the parameters explained above. We here propose a binaural codebook-based method for estimating the STP parameters, and a directional pitch estimator based on the harmonic model for estimating the pitch parameters. The estimated parameters are subsequently used in a binaural speech enhancement framework that is based on the signal model used in \cite{goh1999kalman}. Codebook-based approaches for estimating STP parameters in the single channel case have been previously proposed in \cite{srinivasan2007codebook}, and has been used to estimate the filter parameters required for the Kalman filter for single channel speech enhancement in \cite{mathew2016}. In this work we extend this to the dual channel case, where we assume that there is a wireless link between the HAs. The estimation of STP and pitch parameters using the information on both the left and right channels  leads to a more robust estimation of these parameters. Thus, in this work, we propose a binaural speech enhancement method that is model-based in several ways as 1) the state space equations involved in the Kalman filter takes into account the dynamics of the speech production model; 2) the estimation of STP parameters utilised in the Kalman filter is based on trained spectral models of speech and noise; and 3) the pitch parameters used within the Kalman filter are estimated based on the harmonic model which is a good model for voiced speech. We remark that this paper is an extension of  previous conference papers \cite{mathew2016iwaenc, kavalekalam2017model}. In comparison to \cite{mathew2016iwaenc, kavalekalam2017model}, we have used an improved method for estimating the excitation variances. Moreover, the proposed enhancement framework has been evaluated in more realistic scenarios and subjective listening tests have been conducted to validate the results obtained using objective measures.  

\section{Problem formulation}
\label{sec: binaural_model}
In this section, we formulate the problem and state the assumptions that have been used in this work.
The  noisy signals at the left/right ears at time index $n$ are denoted by
\begin{equation}
\label{eq:left noisy}
z_{l/r}(n) = s_{l/r}(n) + w_{l/r}(n) \hspace{0.7cm} \forall n = 0,1,2\hdots ,
\end{equation}
where $z_{l/r}$, $s_{l/r}$ and $w_{l/r}$ denote the noisy, clean and noise components at the left/right ears, respectively.
It is assumed that the clean speech component is statistically independent with the noise component. Our objective here is to obtain estimates of  the clean speech signals denoted as $\hat{s}_{l/r}(n)$, from the  noisy signals. The processing of the noisy speech using a speech enhancement system to estimate the clean speech signal requires the knowledge of the speech and noise statistics. To obtain this, it is convenient to assume a statsitical model for the speech and noise components, making it easier to estimate the statistics from the noisy signal. In this work, we model the clean speech as an AR process,  which is a  common model used to represent the speech production process \cite{makhoul1975linear}.

We also assume that the speech source is in the nose direction of the listener, so that the clean speech component at the left and right ears can be represented by AR processes having the same parameters,

\begin{equation}
\label{eq:AR_model_speech}
s_{l/r}(n) = \sum\limits_{i=1}^P a_i s_{l/r}(n-i) + u(n),
\end{equation} 
where $\mathbf{a} = [ -a_1, \hdots, -a_P]^T$  is the set of speech AR coefficients,  $P$ is the order of the speech AR process and $u(n)$ is the excitation signal corresponding to the speech signal. Often, $u(n)$ is modelled as white Gaussian noise with variance $\sigma_u^2$  and this will be referred to as the unvoiced (UV) model \cite{gibson1991filtering}. It should be noted that we do not model the reverberation here. Similar to the speech, the noise components are represented by AR processes as,
\begin{equation}
\label{eq:AR_model_noise}
w_{l/r}(n) = \sum\limits_{i=1}^Q c_i w_{l/r}(n-i) + v(n),
\end{equation} 
where $\mathbf{c} = [  -c_1, \hdots, -c_Q]^T$ is the set of noise AR coefficients, $Q$ is the order of the noise AR process and $v(n)$ is  white Gaussian noise with variance $\sigma_v^2$.

As we have seen previously, the excitation signal, $u(n)$, in (\ref{eq:AR_model_speech}) was modelled as a  white Gaussian noise. Although this assumption is suitable for representing unvoiced speech, it is not appropriate for modelling voiced speech. Thus, inspired by \cite{goh1999kalman}, the enhancement framework  here models $u(n)$ as
\begin{equation}
\label{eq: exc_signal_mod}
u(n) = b(p)u(n-p) + d(n),
\end{equation}
where $d(n)$ is white Gaussian noise with variance $\sigma_d^2$, $p$ is the  pitch period and $b(p)\in(0,1)$ is the degree of voicing. In portions containing predominantly voiced speech, $b(p)$ is assumed to be close to $1$ and the variance of $d(n)$ is assumed to be small, whereas in portions of unvoiced speech, $b(p)$ is assumed to be close to zero so that (\ref{eq:AR_model_speech}) simplifies into the conventional unvoiced AR model. The excitation model in (\ref{eq: exc_signal_mod}) when used together with (\ref{eq:AR_model_speech}) is referred to as the voiced-unvoiced (V-UV) model. This model can be easily incorporated into the speech enhancement framework by modifying the state space equations. The incorporation  of the V-UV model into the enhancement framework requires the pitch parameters, $p$ and $b(p)$, in addition to the STP parameters to be estimated from the noisy signal. We would like to remark here that these parameters are usually time varying in the case of speech and noise signals. Herein, these parameters are assumed to be quasi-stationary, and are estimated  for every frame index $f_n =\floor{\frac{n}{M}}+1 $, where $M$ is the frame length. 
The estimation of these parameters will be explained in the subsequent section.

\section{Proposed Enhancement framework}
\label{sec:enh_framework}

\tikzstyle{decision} = [diamond, draw]
\tikzstyle{line} = [draw, -stealth, thick]
\tikzstyle{elli}=[draw, ellipse,minimum height=8mm, text width=5em, text centered]
\tikzstyle{block} = [draw, rectangle,text width=5em, text centered, minimum height=10mm, node distance=10em]
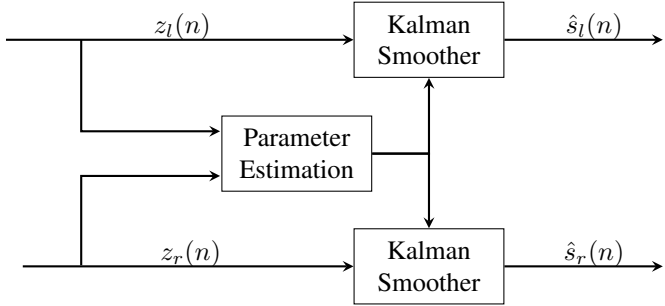
\begin{figure}
\begin{tikzpicture}

\coordinate (start);
\node [block, right of = start,xshift=6em] (kalman) {Kalman Smoother};
\node [block, below of= kalman,xshift=-5em,yshift=2.0cm] (codebook) {Parameter Estimation};
\node [block, below of= kalman,yshift=0.5cm] (kalmanr) {Kalman Smoother};
\coordinate [right of = start](down1);
\coordinate [below of = down1, yshift = -2.0cm](up1);
\coordinate [left of = kalmanr,xshift = -4.4cm](start2);
\coordinate [right of = kalman, xshift = 6em](end);
\coordinate [right of = kalmanr, xshift = 6em](endr);
\coordinate [above of = codebook, yshift = -2em, xshift = -2.9em](codebook1);
\coordinate [below of = codebook, yshift = 2em, xshift = -2.9em](codebook2);

\path [line] (start) -- node[yshift=0.5em, xshift=0.1em] {$z_l(n)$}(kalman);
\path [line] (start2) -- node[yshift=0.5em, xshift=0.1em] {$z_r(n)$}(kalmanr);
\path [line] (codebook) -| node[yshift=0.5em, xshift=4em]{}(kalman); 
\path [line] (down1) |- (codebook1);
\path [line] (up1) |- (codebook2);
\path [line] (codebook) -| (kalmanr);
\path [line] (kalman) -- node[yshift=0.5em, xshift=0.4em] {$\hat{s}_l(n)$}(end);
\path [line] (kalmanr) -- node[yshift=0.5em, xshift=0.4em] {$\hat{s}_r(n)$}(endr);
\end{tikzpicture}
\caption{Basic block diagram of the binaural enhancement framework.}
\label{fig: basic_binaural}
\end{figure}

\subsection{Overview}
\label{ssec: enh_overview}


The enhancement framework proposed here assumes that there is a communication link between the two HAs that makes it possible to exchange information.
 Fig. \ref{fig: basic_binaural} shows the basic block diagram of the proposed enhancement framework. The noisy signals at the left and right ears are enhanced using a fixed lag Kalman smoother (FLKS), which requires the estimation of STP and pitch parameters. These parameters are estimated jointly using the information in the left and right channels. The usage of identical filter parameters at both the ears leads to the preservation of binaural cues.
In this paper, the details regarding the proposed binaural framework will be explained and the performance of the binaural framework will be compared with that of the bilateral framework, where it is assumed that there is no communication link between the two HAs which leads to the filter parameters being  estimated independently at each ear. We will now explain the different components of the proposed enhancement framework in detail.

\subsection{FLKS for speech enhancement}
\label{ssec: kalmam}
As alluded to in the introduction, a  Kalman filter  allows us to take into account the speech production dynamics in the form of state space equations while also accounting for the observation noise. In this work, we use FLKS which is a variant of the Kalman filter. A FLKS gives a better performance than a Kalman filter, but has a higher delay. In this section, we will explain the functioning of FLKS for both the UV and V-UV models that we have introduced in Section \ref{sec: binaural_model}. We assume here that the model parameters are known. For the UV model, the usage of a FLKS (with a smoother delay of $d_s \geq P$) from a speech enhancement perspective requires the AR signal model in (\ref{eq:AR_model_speech}) to be written as a state space form as shown below

\begin{equation}
\label{eq:speech_state_space_UV}
\bar{\mathbf{s}}_{l/r}(n) = \mathbf{A}(f_n)\bar{\mathbf{s}}_{l/r}(n-1)+\mathbf{\Gamma}_1 u(n),
\end{equation}
where $\bar{\mathbf{s}}_{l/r}(n) = [s_{l/r}(n), s_{l/r}(n-1), \hdots ,s_{l/r}(n-d_{s})]^T$ is the state vector containing the $d_{s}+1$ recent speech samples, $\mathbf{\Gamma}_1=[1, 0, \hdots ,0]^T$ is a $(d_s+1)\times 1$ vector, $u(n) = d(n)$ and $\mathbf{A}(f_n)$ is the $(d_s+1) \times (d_s+1)$ speech state transition matrix written as
\begin{equation}
\mathbf{A}(f_n)=\begin{bmatrix}
-\mathbf{a}(f_n)^T & \mathbf{0}^T & 0 \\
\mathbf{I}_P & \mathbf{0} &  \mathbf{0} \\
\mathbf{0} & \mathbf{I}_{d_s-P} & \mathbf{0}
\end{bmatrix}.
\end{equation}
The state space equation for the noise signal in (\ref{eq:AR_model_noise}) is similarly written as
\begin{equation}
\label{eq:noise state space}
\bar{\mathbf{w}}_{l/r}(n) = \mathbf{C}(f_n)\bar{\mathbf{w}}_{l/r}(n-1)+\mathbf{\Gamma}_2 v(n),
\end{equation}
where $\bar{\mathbf{w}}_{l/r}(n) = [w_{l/r}(n), w_{l/r}(n-1), \hdots ,w_{l/r}(n-Q+1)]^T$, $\mathbf{\Gamma}_2=[1, 0, \hdots, 0]^T$ is a $Q\times 1$ vector and
\begin{equation}
\mathbf{C}(f_n)=\begin{bmatrix}
 \left[ c_1(f_n), \hdots, c_{Q-1}(f_n)\right] & c_Q(f_n)\\
  \mathbf{I}_{Q-1}   & \mathbf{0} \\
\end{bmatrix}
\end{equation}
 is a $Q \times Q$ matrix. The state space equations in (\ref{eq:speech_state_space_UV}) and (\ref{eq:noise state space}) are combined to form a concatenated state space equation for the UV model as
\begin{equation*}
\scalebox{0.95}{$
\label{eq: conc_statespace}
\begin{bmatrix}
\bar{\mathbf{s}}_{l/r}(n) \\
\bar{\mathbf{w}}_{l/r}(n)
\end{bmatrix} = \begin{bmatrix}
\mathbf{A}(f_n) & \mathbf{0} \\
\mathbf{0} & \mathbf{C}(f_n)
\end{bmatrix}\begin{bmatrix}
\bar{\mathbf{s}}_{l/r}(n-1) \\
\bar{\mathbf{w}}_{l/r}(n-1) 
\end{bmatrix} + \begin{bmatrix}
\mathbf{\Gamma}_1 & \mathbf{0} \\
\mathbf{0} & \mathbf{\Gamma}_2
\end{bmatrix}\begin{bmatrix}
d(n) \\ v(n)
\end{bmatrix}
$}
\end{equation*}
which can be rewritten as
\begin{equation}
\label{eq: final state space_UV}
\bar{\mathbf{x}}^{\text{\tiny{UV}}}_{l/r}(n) \triangleq \mathbf{F}^{\text{\tiny{UV}}}(f_n)\mathbf{x}(n-1)+\mathbf{\Gamma}_3 \mathbf{y}(n),
\end{equation}
where $\bar{\mathbf{x}}^{\text{\tiny{UV}}}_{l/r}(n) = \left[\bar{\mathbf{s}}_{l/r}(n)^T \, \bar{\mathbf{w}}_{l/r}(n)^T   \right]^T$ is the concatenated state space vector and $\mathbf{F}^{\text{\tiny{UV}}}(f_n)$ is the concatenated state transition matrix for the UV model.
The observation equation to obtain the noisy signal is then written as
\begin{equation}
\label{eq: final observation_UV}
z_{l/r}(n) = \mathbf{\Gamma}^{{\text{\tiny{UV}}}^T} \bar{\mathbf{x}}^{\text{\tiny{UV}}}_{l/r}(n),
\end{equation}
where $\mathbf{\Gamma}^{{\text{\tiny{UV}}}} = \left[ \mathbf{\Gamma}_1^T\, \mathbf{\Gamma}_2^T \right]^T$.
The state space equation (\ref{eq: final state space_UV}) and the observation equation (\ref{eq: final observation_UV}) can then be used to formulate the prediction and correction stages of the FLKS for the UV model.
We will now explain the formulation of the state space equations for the V-UV model.
The state space equation for the V-UV model of speech is written as
\begin{equation}
\label{eq:speech_state_space}
\bar{\mathbf{s}}_{l/r}(n) = \mathbf{A}(f_n)\bar{\mathbf{s}}_{l/r}(n-1)+\mathbf{\Gamma}_1 u(n),
\end{equation}
where the excitation signal in (\ref{eq: exc_signal_mod}) is also modelled as a state space equation as
\begin{equation}
\label{eq:exc_state_space}
\bar{\mathbf{u}}(n) = \mathbf{B}(f_n)\bar{\mathbf{u}}(n-1)+\mathbf{\Gamma}_4 d(n),
\end{equation}
where $\bar{\mathbf{u}}(n) = [u(n),\, u(n-1), \hdots, u(n-p_{\text{max}}+1)]^T$, $p_{\text{max}}$ is the maximum pitch period in integer samples, $\mathbf{\Gamma}_4=[1, 0 \hdots 0]^T$ is a $(p_{\text{max}})\times 1$ vector and
\begin{equation}
\mathbf{B}(f_n)=\begin{bmatrix}
 \left[ b(1), \hdots, b(p_{\text{max}}-1)\right] & b(p_{\text{max}})\\
  \mathbf{I}_{p_{\text{max}}-1}   & \mathbf{0} \\
\end{bmatrix}
\end{equation}
is a $p_{\text{max}}\times p_{\text{max}}$ matrix  where $b(i) = 0 \,\,\forall i \neq p({f_n})$. The concatenated state space equation for the V-UV model is 
\begin{equation*}
\begin{split}
\begin{bmatrix}
\bar{\mathbf{s}}_{l/r}(n) \\
\mathbf{u}(n+1) \\
\bar{\mathbf{w}}_{l/r}(n)
\end{bmatrix} =  \begin{bmatrix}
\mathbf{A}(f_n) & \mathbf{\Gamma}_1\mathbf{\Gamma}_2^T &\mathbf{0} \\
\mathbf{0} & \mathbf{B}(f_n) & \mathbf{0}\\
\mathbf{0}& \mathbf{0} & \mathbf{C}(f_n)
\end{bmatrix}
\begin{bmatrix}
\bar{\mathbf{s}}_{l/r}(n-1) \\
\bar{\mathbf{u}}(n) \\
\bar{\mathbf{w}}_{l/r}(n-1)
\end{bmatrix}\\
+ \begin{bmatrix}
\mathbf{0} & \mathbf{0} \\
\mathbf{\Gamma}_4 & \mathbf{0} \\
\mathbf{0} & \mathbf{\Gamma}_2
\end{bmatrix}\begin{bmatrix}
d(n+1)\\ v(n)
\end{bmatrix},
\end{split}
\end{equation*}
which can also be written as
\begin{equation}
\label{eq: final_state_space_VUV}
\bar{\mathbf{x}}^{\text{\tiny{V-UV}}}_{l/r}(n+1) \triangleq \mathbf{F}^{\text{\tiny{V-UV}}}(f_n)\bar{\mathbf{x}}^{\text{\tiny{V-UV}}}_{l/r}(n)+\mathbf{\Gamma}_5 \mathbf{g}(n+1),
\end{equation}
where $\bar{\mathbf{x}}^{\text{\tiny{V-UV}}}_{l/r}(n+1)= [\bar{\mathbf{s}}_{l/r}(n)^T \,  \bar{\mathbf{u}}(n+1)^T \, \bar{\mathbf{w}}_{l/r}(n)^T]^T$ is the concatenated state space vector, $\mathbf{g}(n+1) = [d(n+1)\, v(n)]^T$ and $\mathbf{F}^{\text{\tiny{V-UV}}}(f_n)$ is the concatenated state transition matrix for the V-UV model.
 The observation equation to obtain the noisy signal is written as
\begin{equation}
\label{eq: final observation_V_UV}
z_{l/r}(n) = \mathbf{\Gamma}^{{\text{\tiny{V-UV}}}^T} \bar{\mathbf{x}}^{\text{\tiny{V-UV}}}_{l/r}(n+1),
\end{equation}
where $\mathbf{\Gamma}^{{\text{\tiny{V-UV}}}} = \left[\mathbf{\Gamma}_1^T\, \mathbf{0}^T\, \mathbf{\Gamma}_2^T \right]^T$. The state space equation (\ref{eq: final_state_space_VUV}) and the observation equation (\ref{eq: final observation_V_UV}) can then be used to formulate the prediction and correction stages of the FLKS for the V-UV model (see Appendix \ref{ssec:App_pred}). It can be seen that the formulation of the prediction and correction stages of the FLKS  requires the knowledge of the speech and noise STP parameters, and the clean speech pitch parameters. The estimation of these model parameters are explained in the subsequent sections.

\subsection{Codebook-based binaural estimation of STP parameters}
\label{ssec : codebook}
As mentioned in the introduction, the estimation of the speech and noise STP parameters forms a very critical part of the proposed enhancement framework. 
These parameters are here estimated using a codebook-based approach. The estimation of STP parameters using a codebook-based approach, when having access to a single channel noisy signal has been previously proposed in \cite{srinivasan2007codebook,he2017multiplicative}. Here, we extend this to the case when we have access to binaural noisy signals. Codebook-based estimation of STP parameters uses the a priori information about speech and noise spectral shapes stored in trained speech and noise codebooks in the form of speech and noise AR coefficients respectively. The codebooks offer us an elegant way of including prior information about the speech and noise spectral models e.g. if the enhancement system present in the HA has to operate in a particular noisy environment, or mainly process speech from a particular set of speakers, the codebooks can be trained accordingly. Contrarily, if we do not have any specific information regarding the speaker or the noisy environment, we can still train general codebooks from a large database consisting of different speakers and noise types. We would like to remark here that we  assume the UV model of speech for the estimation of STP parameters.

A Bayesian framework is utilised to estimate the parameters for every frame index.
 Thus, the random variables (r.v.) corresponding to the parameters to be estimated for the $f_n^{\text{th}}$ frame are concatenated to form a single vector $\boldsymbol{\theta}(f_n) = [\boldsymbol{\theta}_s(f_n)^T \,\,  \boldsymbol{\theta}_w(f_n)^T]^T = [\mathbf{a}(f_n)^T \, \sigma_d^2(f_n) \,\mathbf{c}(f_n)^T \, \sigma_v^2(f_n)]^T$, where $\mathbf{a}(f_n)$ and $\mathbf{c}(f_n)$ are r.v. representing the speech and noise AR coefficients, and $\sigma_d^2(f_n)$ and $\sigma_v^2(f_n)$ are r.v. representing the speech and noise excitation variances. The MMSE estimate of the parameter vector is 
\begin{equation}
\label{eq:cont MMSE parameter est1}
\hat{\boldsymbol{\theta}}(f_n) = \mathbb{E}(\boldsymbol{\theta}(f_n)|\mathbf{z}_l(f_nM),\mathbf{z}_r(f_nM)), 
\end{equation}
where $\mathbb{E}(\cdot)$ is the expectation operator and $\mathbf{z}_{l/r}(f_nM) = \left[ z_{l/r}(f_nM), \hdots, z_{l/r}(f_nM+m),  \hdots,z_{l/r}(f_nM+M-1)\right]^T$ denotes the $f_n^{\text{th}}$  frame  of noisy speech at the left/right ears. The frame index, $f_n$, will be left out for the remainder of the section for notational convenience. Equation (\ref{eq:cont MMSE parameter est1}) is then rewritten as
\begin{equation}
\label{eq:cont MMSE parameter est}
\hat{\boldsymbol{\theta}} = \int_\Theta \boldsymbol{\theta}\frac{p(\mathbf{z}_l,\mathbf{z}_r|\boldsymbol{\theta})\,\,p(\boldsymbol{\theta})}{p(\mathbf{z}_l,\mathbf{z}_r)}d\boldsymbol{\theta},
\end{equation}
where $\Theta$ denotes the combined support space of the parameters to be estimated. Since we assumed that the speech and noise are independent (see Section \ref{sec: binaural_model}), it follows that $p(\boldsymbol{\theta}) = p(\boldsymbol{\theta}_s)p(\boldsymbol{\theta}_w)$ where $\boldsymbol{\theta}_s$ and $\boldsymbol{\theta}_w$ speech and noise STP parameters respectively. Furthermore, the speech and noise AR coefficients are assumed to be independent with the excitation  variances leading to $p(\boldsymbol{\theta}_s) = p(\mathbf{a})p(\sigma_d^2)$ and $p(\boldsymbol{\theta}_w) = p(\mathbf{c})p(\sigma_v^2)$. Using the aforementioned assumptions, (\ref{eq:cont MMSE parameter est}) is rewritten as
\begin{equation}
\label{eq: MMSE_2}
\hat{\boldsymbol{\theta}} = \int_\Theta \boldsymbol{\theta}\frac{p(\mathbf{z}_l,\mathbf{z}_r|\boldsymbol{\theta})\, p(\mathbf{a}) p(\sigma_d^2) p(\mathbf{c}) p(\sigma_v^2) }{p(\mathbf{z}_l,\mathbf{z}_r)}d\boldsymbol{\theta}.
\end{equation}
The probability density of the AR coefficients is here modelled as a sum of Dirac delta functions centered around each codebook entry as $p(\mathbf{a}) = \frac{1}{N_s}\sum_{i=1}^{N_s} \delta(\mathbf{a}-\mathbf{a}_i) $ and $p(\mathbf{c}) = \frac{1}{N_w}\sum_{j=1}^{N_w} \delta(\mathbf{c}-\mathbf{c}_j)$, where $\mathbf{a}_i$ is the $i^{th}$ entry of the speech codebook (of size $N_s$), $\mathbf{c}_j$ is the $j^{th}$ entry of the noise codebook (of size $N_w$) . Defining $\boldsymbol{\theta}_{ij} \triangleq [\mathbf{a}_i^T \, \sigma_{d}^{2} \, \mathbf{c}_j^T \, \sigma_{v}^{2}]^T$,  (\ref{eq: MMSE_2}) can be rewritten as
\begin{equation}
\label{eq: MMSE_3}
\hat{\boldsymbol{\theta}} \!= \!\!\frac{1}{N_sN_w}\sum\limits_{i=1}^{N_s}\sum\limits_{j=1}^{N_w} \!\!\int_{\sigma_d^2} \int_{\sigma_v^2}\!\!\!\! \boldsymbol{\theta}_{ij}\frac{p(\mathbf{z}_l,\mathbf{z}_r|\boldsymbol{\theta}_{ij})\,  p(\sigma_d^2) p(\sigma_v^2) }{p(\mathbf{z}_l,\mathbf{z}_r)}d\sigma_d^2d\sigma_v^2.
\end{equation}
For a particular set of speech and noise AR coefficients, $\mathbf{a}_i$ and $\mathbf{c}_j$, it can be shown that the likelihood, $p(\mathbf{z}_l,\mathbf{z}_r|\boldsymbol{\theta}_{ij})$, decays rapidly from its maximum value when there is a small deviation in the excitation variances from its true value \cite{srinivasan2007codebook} (see Appendix \ref{ssec:behav}). If we then approximate the true values of the excitation variances with the corresponding  maximum likelihood (ML) estimates  denoted as $\sigma_{d,ij}^{2}$ and $\sigma_{v,ij}^{2} $, the likelihood term $p(\mathbf{z}_l,\mathbf{z}_r|\boldsymbol{\theta}_{ij})$ can be approximated as $p(\mathbf{z}_l,\mathbf{z}_r|\boldsymbol{\theta}_{ij}) \delta(\sigma_{d}^2-\sigma_{d,ij}^{2} ) \delta(\sigma_{v}^2-\sigma_{v,ij}^{2})$. Defining $\boldsymbol{\theta}_{ij}^{\text{ML}} \triangleq [\mathbf{a}_i^T \, \sigma_{d,ij}^{2} \, \mathbf{c}_j^T \, \sigma_{v,ij}^{2}]^T$, and using the above approximation and the property, $\int_x f(x) \delta(x-x_0) dx = f(x_0)$, we can rewrite (\ref{eq: MMSE_3}) as

\begin{equation}
\label{eq: final_discrete_MMSE}
\!\hat{\boldsymbol{\theta}} = \frac{1}{N_sN_w}\sum\limits_{i=1}^{N_s}\sum\limits_{j=1}^{N_w} \boldsymbol{\theta}_{ij}^{\text{ML}}\frac{p(\mathbf{z}_l,\mathbf{z}_r|\boldsymbol{\theta}_{ij}^{\text{ML}}) p(\sigma_{d,ij}^{2})p(\sigma_{v,ij}^{2})}{p(\mathbf{z}_l,\mathbf{z}_r)},
\end{equation}
where
\begin{equation*}
p(\mathbf{z}_l, \mathbf{z}_r) = \frac{1}{N_sN_w}\sum\limits_{i=1}^{N_s}\sum\limits_{j=1}^{N_w} p(\mathbf{z}_l,\mathbf{z}_r|\boldsymbol{\theta}_{ij}^{\text{ML}}) p(\sigma_{d,ij}^{2})p(\sigma_{v,ij}^{2}).
\end{equation*}
Details regarding the prior distributions used for the excitation variances is given in Appendix \ref{ssec: gamma}. 
It can be seen from (\ref{eq: final_discrete_MMSE}) that the final estimate of the parameter vector is a weighted linear combination of $\boldsymbol{\theta}_{ij}^{\text{ML}}$ with weights proportional to $p(\mathbf{z}_l,\mathbf{z}_r|\boldsymbol{\theta}_{ij}^{\text{ML}}) p(\sigma_{d,ij}^{2})p(\sigma_{v,ij}^{2})$. To compute this, we need to first obtain the ML estimates of the excitation variances for a given set of speech and noise AR coefficients, $\mathbf{a}_i$ and $\mathbf{c}_j$, as
\begin{equation}
\label{eq: opt_1}
\{\sigma_{d,ij}^{2}, \sigma_{v,ij}^{2} \}= \argmax_{\sigma_{d}^{2}, \sigma_{v}^{2}\geq 0} \,\,\,\,p(\mathbf{z}_l,\mathbf{z}_r|\boldsymbol{\theta}_{ij}).
\end{equation}
For the models we have assumed previously in Section \ref{sec: binaural_model}, we can show that $\mathbf{z}_l$ and $\mathbf{z}_r$ are statistically independent given $\boldsymbol{\theta}_{ij}$ \cite[Sec 8.2.2]{christopher2016pattern}, which results in 
\begin{equation*}
p(\mathbf{z}_l,\mathbf{z}_r|\boldsymbol{\theta}_{ij}) = p(\mathbf{z}_l|\boldsymbol{\theta}_{ij}) p(\mathbf{z}_r|\boldsymbol{\theta}_{ij}).
\end{equation*}
We first derive the likelihood for the left channel, $p(\mathbf{z}_l|\boldsymbol{\theta}_{ij})$, using the assumptions we have introduced previously in Section \ref{sec: binaural_model}. Using these assumptions, frame of speech and noise component associated with the noisy frame $\mathbf{z}_l$ denoted  by $\mathbf{s}_{l} $ and  $\mathbf{w}_{l} $ respectively can be expressed as
\begin{eqnarray*}
p(\mathbf{s}_{l}|\sigma_d^2, \mathbf{a}_i) &\sim \mathcal{N}(\mathbf{0},\, \sigma_d^2 \mathbf{R}_s(\mathbf{a}_i)) 
\\
p(\mathbf{w}_{l}|\sigma_v^2, \mathbf{c}_j) &\sim \mathcal{N}(\mathbf{0},\, \sigma_v^2 \mathbf{R}_w(\mathbf{c}_j)),
\end{eqnarray*}
where  $\mathbf{R}_s(\mathbf{a}_i)$ is the normalised speech covariance matrix and $\mathbf{R}_w(\mathbf{c}_j)$ is the normalised noise covariance matrix. These matrices can be asymptotically approximated as circulant matrices which can be diagonalised using the Fourier transform as \cite{srinivasan2007codebook,gray2006toeplitz},
\begin{equation*}
\label{eq:Rs_diagn}
\mathbf{R}_s(\mathbf{a}_i) =  \mathbf{F}\mathbf{D}_{s_{i}}\mathbf{F}^H \,\,\,\,\text{and}\,\,\,\, \mathbf{R}_w(\mathbf{c}_j) = \mathbf{F}\mathbf{D}_{w_j}\mathbf{F}^H,
\end{equation*}
where $\mathbf{F}$ is the discrete Fourier transform (DFT) matrix defined as
$
[\mathbf{F}]_{m,k} =\frac{1}{\sqrt{M}} \exp(\frac{\imath 2\pi mk}{M}), \,\,\,\,\forall m,k = 0, \hdots M-1
$ where $k$ represents the frequency index and
\begin{equation*} 
\mathbf{D}_{s_i} = (\mathbf{\Lambda}_{s_i}^H\mathbf{\Lambda}_{s_i})^{-1},\,\,\,\,\,\,\,\,\,  \mathbf{\Lambda}_{s_i} = \text{diag}\left(\sqrt{M}\mathbf{F}^H \begin{bmatrix}
1\\
\mathbf{a}_i \\
\mathbf{0}
\end{bmatrix}\right),
\end{equation*}
\begin{equation*} 
\mathbf{D}_{w_j} = (\mathbf{\Lambda}_{w_j}^H\mathbf{\Lambda}_{w_j})^{-1},\,\,\,\,\,\,\,\,\,  \mathbf{\Lambda}_{w_j} = \text{diag}\left(\sqrt{M}\mathbf{F}^H \begin{bmatrix}
1\\
\mathbf{c}_j \\
\mathbf{0}
\end{bmatrix}\right).
\end{equation*}
Thus we obtain the likelihood for the left channel as,
\begin{equation*}
p(\mathbf{z}_l|\boldsymbol{\theta}_{ij}) \sim \mathcal{N}(\mathbf{0},\,  \sigma_{d}^{2}\mathbf{F}\mathbf{D}_{s_{i}}\mathbf{F}^H + \sigma_{v}^{2}\mathbf{F}\mathbf{D}_{w_{j}}\mathbf{F}^H).
\end{equation*}
The log-likelihood $\text{ln}p(\mathbf{z}_l|\boldsymbol{\theta}_{ij})$ is then given by
\begin{equation}
\label{eq:log_l_D}
\begin{split}
\text{ln}p(\mathbf{z}_l|\boldsymbol{\theta}_{ij})\,\, \myeq \,\, \text{ln}\Big{|} \sigma_{d}^{2}\mathbf{F}\mathbf{D}_{s_i}\mathbf{F}^H + \sigma_{v}^{2}\mathbf{F}\mathbf{D}_{w_j}\mathbf{F}^H\Big{|}^{-\frac{1}{2}}\\-\frac{1}{2} \mathbf{z}_l^T \left[ \sigma_{d}^{2}\mathbf{F}\mathbf{D}_{s_i}\mathbf{F}^H + \sigma_{v}^{2}\mathbf{F}\mathbf{D}_{w_j}\mathbf{F}^H \right]^{-1}\mathbf{z}_l,
\end{split}
\end{equation}
where $ \myeq$ denotes equality up to a constant and $|\cdot|$ denotes the matrix determinant operator.
Denoting $\frac{1}{A_s^i(k)}$ as the $k^{\text{th}}$ diagonal element of $\mathbf{D}_{s_i}$ and $\frac{1}{A_w^i(k)}$ as the $k^{\text{th}}$ diagonal element of $\mathbf{D}_{w_j}$, (\ref{eq:log_l_D}) can be rewritten as 
\begin{equation}
\label{eq: log_l_D_2}
\begin{split}
&\text{ln}p(\mathbf{z}_l|\boldsymbol{\theta}_{ij}) \,\, \myeq \,\,  \text{ln} \prod_{k=0}^{K-1} \left(  \frac{\sigma_{d}^{2}}{A_s^i(k)} + \frac{\sigma_{v}^{2}}{A_w^j(k)} \right)^{-\frac{1}{2}} \\ & -\frac{1}{2} \mathbf{z}_l^T \mathbf{F}  \begin{bmatrix} 
\frac{\sigma_{d}^{2}}{A_s^i(0)} + \frac{\sigma_{v}^{2}}{A_w^j(0)} & \mathbf{0} &0 \\
\mathbf{0} & \ddots & \mathbf{0} \\
0 & \mathbf{0} & \frac{\sigma_{d}^{2}}{A_s^i(K-1)} + \frac{\sigma_{v}^{2}}{A_w^j(K-1)} 
\end{bmatrix}^{-1}\!\!\!\!\!\!\!\!  \mathbf{F}^H \mathbf{z}_l.
 \end{split}
\end{equation}
Defining the modelled spectrum as
$
\hat{P}_{z_{ij}}(k) \triangleq \frac{\sigma_{d}^{2}}{A_s^i(k)} + \frac{\sigma_{v}^{2}}{A_w^j(k)},
$
 (\ref{eq: log_l_D_2}) can be written as
\begin{equation}
\begin{split}
\text{ln}p(\mathbf{z}_l|\boldsymbol{\theta}_{ij}) \,\, \myeq \,\, \text{ln}\prod_{k=0}^{K-1} \left(  \hat{P}_{z_{ij}}(k) \right)^{-\frac{1}{2}}  - \frac{1}{2}  \sum\limits_{k=0}^{K-1} \frac{P_{z_l}(k)}{\hat{P}_{z_{ij}}(k)},
\end{split}
\end{equation}
where  $P_{z_l}(k)$ is the squared magnitude of the $k^{\text{th}}$ element of the vector $\mathbf{F}^H \mathbf{z}_l$. Thus, 
\begin{equation}
\text{ln}p(\mathbf{z}_l|\boldsymbol{\theta}_{ij}) \,\, \myeq \,\,  - \frac{1}{2}\sum_{k=0}^{K-1} \left(\frac{P_{z_l}(k)}{\hat{P}_{z_{ij}}(k)} + \text{ln} \hat{P}_{z_{ij}}(k)\right).
\end{equation}
We can then see that the log-likelihood is  equal, up to a constant, to the Itakura-Saito (IS) divergence between $\mathbf{P}_{z_l}$ and $\hat{\mathbf{P}}_{z_{ij}}$ which is defined as \cite{itakura1968analysis} $$d_{\mathrm{IS}}(\mathbf{P}_{z_l},\hat{\mathbf{P}}_{z_{ij}}) = \frac{1}{K}\sum_{k=0}^{K-1} \left(\frac{P_{z_l}(k)}{\hat{P}_{z_{ij}}(k)} - \text{ln} \frac{P_{z_l}(k)}{\hat{P}_{z_{ij}}(k)} - 1\right), $$
where $\mathbf{P}_{z_l} = \left[ P_{z_l}(0), \hdots, P_{z_l}(K-1) \right]^T$ and $\hat{\mathbf{P}}_{z_{ij}} = \left[ \hat{P}_{z_{ij}}(0), \hdots, \hat{P}_{z_{ij}}(K-1) \right]^T$.
 Using the same result for the right ear, the optimisation problem in (\ref{eq: opt_1}), under the aforementioned conditions can be equivalently written as
\begin{equation}
\{\sigma_{d,ij}^{2}, \sigma_{v,ij}^{2}\}\!=\! \argmin_{\sigma_{d}^{2}, \sigma_{v}^{2}\geq 0}\!\! \,\,\left[ d_{\mathrm{IS}}(\mathbf{P}_{z_l},\hat{\mathbf{P}}_{z_{ij}}) \!+\! d_{\mathrm{IS}}(\mathbf{P}_{z_r},\!\hat{\mathbf{P}}_{z_{ij}})\right]\!\!.
\label{eq:opt_total_IS}
\end{equation}
Unfortunately, it is not possible to get a closed form expression for the excitation variances by minimising (\ref{eq:opt_total_IS}). Instead, this is solved iteratively using the multiplicative update (MU) method \cite{lee2001algorithms}. 
 For notational convenience, $\hat{\mathbf{P}}_{z_{ij}}$ can be written as
$
\hat{\mathbf{P}}_{z_{ij}} = \mathbf{P}_{s,i} \sigma_{d}^{2} + \mathbf{P}_{w,j} \sigma_{v}^{2},
\label{eq: mod_vec }
$
where
\begin{equation*}
\scalebox{0.95}{$
\mathbf{P}_{s,i} = \left[ \frac{1}{A_s^i(0)}, \hdots ,\frac{1}{A_s^i(K-1)} \right]^T,\,\,\,\,
\mathbf{P}_{w,j} = \left[ \frac{1}{A_w^j(0)}, \hdots, \frac{1}{A_w^j(K-1)} \right]^T$}.
\end{equation*}
Defining $\mathbf{P}_{ij} = \left[ \mathbf{P}_{s,i} \,\,\,\mathbf{P}_{w,j} \right]$, and $\mathbf{\Sigma}_{ij}^{(l)} = [\sigma_{d,ij}^{2(l)} \,\, \sigma_{v,ij}^{2(l)}]^T$ where $\sigma_{d,ij}^{2(l)}$ and $\sigma_{v,ij}^{2(l)}$ represents the ML estimates of the excitation variances at the $l^{\text{th}}$ MU  iteration, the values for the excitation variances using the MU method are computed iteratively as \cite{fevotte2009nonnegative},
\begin{equation}
\label{eq:exc_var_ML_1}
\sigma_{d,ij}^{2(l+1)} \leftarrow \sigma_{d,ij}^{2(l)} \frac{\mathbf{P}_{s,i}^T \left[ (\mathbf{P}_{ij} \mathbf{\Sigma}_{ij}^{(l)})^{-2} \cdot (\mathbf{P}_{z_l} + \mathbf{P}_{z_r})\right]}{2 \mathbf{P}_{s,i}^T(\mathbf{P}_{ij} \mathbf{\Sigma}_{ij}^{(l)})^{-1}} ,
\end{equation}
\begin{equation}
\label{eq:exc_var_ML_2}
\sigma_{v,ij}^{2(l+1)} \leftarrow \sigma_{v,ij}^{2(l)} \frac{\mathbf{P}_{w,j}^T \left[ (\mathbf{P}_{ij} \mathbf{\Sigma}_{ij}^{(l)})^{-2}\cdot (\mathbf{P}_{z_l} + \mathbf{P}_{z_r})\right]}{2 \mathbf{P}_{w,j}^T(\mathbf{P}_{ij} \mathbf{\Sigma}_{ij}^{(l)})^{-1}} ,
\end{equation}
where ($\cdot$) denotes the element wise multiplication operator and $(\cdot)^{-2}$ denotes element-wise inverse squared operator.
The excitation variances estimated using (\ref{eq:exc_var_ML_1}) and (\ref{eq:exc_var_ML_2}) lead to the minimisation of the cost function in (\ref{eq:opt_total_IS}). 
Using these results, $p(\mathbf{z}_l,\mathbf{z}_r|\boldsymbol{\theta}_{ij}^{\text{ML}})$ can be written as
\begin{equation}
\label{eq:final weight}
p(\mathbf{z}_l,\mathbf{z}_r|\boldsymbol{\theta}_{ij}^{\text{ML}}) = Ce^{\left(-\frac{M}{2}\left[ d_{\mathrm{IS}}(\mathbf{P}_{z_l},\hat{\mathbf{P}}_{z_{ij}}^{\text{ML}}) + d_{\mathrm{IS}}(\mathbf{P}_{z_r},\hat{\mathbf{P}}_{z_{ij}}^{\text{ML}})\right]\right)},
\end{equation}
where $C$ is a normalisation constant,
and $\hat{\mathbf{P}}_{z_{ij}}^{\text{ML}} = [\hat{P}_{z_{ij}}^{\text{ML}}(0), \hdots ,\hat{P}_{z_{ij}}^{\text{ML}}(K-1) ]^T$ and
\begin{equation}
\label{eq:modelled_spectrum}
\hat{P}_{z_{ij}}^\text{{ML}}(k) = \frac{\sigma_{d,ij}^{2}}{A_s^i(k)} + \frac{\sigma_{v,ij}^{2}}{A_w^j(k)}.
\end{equation}
Once the likelihoods are calculated using (\ref{eq:final weight}), they are substituted into (\ref{eq: final_discrete_MMSE}) to get the final estimate of the speech and noise STP parameters. 
 Some other practicalities involved in the estimation procedure of the STP parameters are explained next.
 
\pgfplotsset{every axis/.append style={
                   legend style={font=\scriptsize,line width=0.8pt,mark size=1.8pt},
  }}

%

\subsubsection{Adaptive noise codebook}
\label{ssec: Dual noise psd}
The noise codebook used for the estimation of the STP parameters is usually generated by using a training sample consisting of the noise type of interest. However, there might be scenarios where the noise type is not known a priori. In such scenarios, to make the enhancement system more robust, the noise codebook can be appended with an entry corresponding to the noise power spectral density (PSD) estimated using another dual channel method. Here, we utilise such a dual channel method for estimating the noise PSD \cite{dorbecker1996combination}, which requires the transmission of noisy signals between the HAs.
 The estimated dual channel noise PSD, $\hat{P}^{\text{DC}}_w (k)$, is then used to find the AR coefficients and the variance representing the noise spectral envelope. At first, the autocorrelation coefficients corresponding to the noise PSD estimate are computed using the Wiener-Khinchin theorem  as
\begin{equation*}
r_{ww}(q) = \sum\limits_{k=0}^{K-1} \hat{P}^{\text{DC}}_w (k) \exp\left(\imath 2\pi \frac{qk}{K}\right), \,\,\, 0\leq q \leq Q.
\end{equation*}
Subsequently, the AR coefficients denoted by $\hat{\mathbf{c}}^{\text{DC}} = [1,\, \hat{c}^{\text{DC}}_1, \hdots, \hat{c}^{\text{DC}}_Q]^T$, and the excitation variance corresponding to the dual channel noise PSD estimate are  estimated by Levinson-Durbin recursive algorithm \cite[p. 100]{stoica2005spectral}.
The estimated AR coefficient vector, $\hat{\mathbf{c}}^{\text{DC}}$, is then appended to the noise codebook. The final estimate of the noise excitation variance can be taken as a mean of variance obtained from the dual channel estimate and the variance obtained from (\ref{eq: final_discrete_MMSE}).
 It should be noted that, in the case a noise codebook is not available a priori, the speech codebook can be used in conjunction with dual channel noise PSD estimate alone. This leads to a reduction in the computational complexity. Some other dual channel noise PSD estimation algorithms present in the literature are \cite{kamkar2009improved, jeub2011robust}, and these can in principle also be included in the noise codebook.

\subsection{Directional pitch estimator}
\label{ssec: pitch_estimator}
\begin{figure}[t] \centering  \newlength\figureheight \newlength\figurewidth\setlength\figureheight{3.0cm} \setlength\figurewidth{4.5cm} 
%
%
\definecolor{mycolor1}{rgb}{0.00000,0.44700,0.74100}%
\begin{tikzpicture}

\begin{axis}[%
width=0.95092\figurewidth,
height=0.85\figureheight,
at={(0\textwidth,0\textwidth)},
scale only axis,
xmin=10.6451612903226,
xmax=88.8940092165899,
xlabel = Frame index,
ylabel = Hz,
xmajorgrids,
ymin=81.5102040816327,
ymax=347.469387755102,
ymajorgrids
]
\addplot [color=mycolor1,only marks,mark=asterisk,mark options={solid},forget plot]
  table[row sep=crcr]{%
1	286.5\\
2	181.5\\
3	190\\
4	188\\
5	295\\
6	232\\
7	264.5\\
8	182.5\\
9	178.5\\
10	166.5\\
11	172.5\\
12	179.5\\
13	181.5\\
14	186\\
15	191.5\\
16	193\\
17	199.5\\
18	209.5\\
19	218\\
20	233.5\\
21	251.5\\
22	256\\
23	253\\
24	119\\
25	118\\
26	115\\
27	118\\
28	236\\
29	236.5\\
30	240.5\\
31	240\\
32	222.5\\
33	213\\
34	203\\
35	206.5\\
36	215\\
37	224\\
38	230\\
39	232.5\\
40	230\\
41	217\\
42	217.5\\
43	210\\
44	205.5\\
45	202\\
46	199\\
47	192\\
48	190\\
49	197\\
50	206\\
51	210\\
52	221\\
53	235\\
54	251.5\\
55	103.5\\
56	303\\
57	305\\
58	306.5\\
59	258\\
60	222.5\\
61	186.5\\
62	181.5\\
63	171.5\\
64	168.5\\
65	162\\
66	174\\
67	178\\
68	174\\
69	170\\
70	167\\
71	164\\
72	161\\
73	159.5\\
74	157.5\\
75	155.5\\
76	153.5\\
77	151\\
78	138.5\\
79	142\\
80	151\\
81	154\\
82	149.5\\
83	143\\
84	141\\
85	141.5\\
86	143\\
87	144.5\\
88	144.5\\
89	143.5\\
90	142\\
91	142.5\\
92	379\\
93	166\\
94	176.5\\
95	193\\
96	245.5\\
97	309.5\\
98	112\\
99	135.5\\
100	301.5\\
101	126\\
102	376.5\\
};
\addplot [color=red,solid,forget plot]
  table[row sep=crcr]{%
1	132.8125\\
2	179.6875\\
3	99.609375\\
4	173.828125\\
5	166.015625\\
6	171.875\\
7	187.5\\
8	162.109375\\
9	167.96875\\
10	167.96875\\
11	173.828125\\
12	179.6875\\
13	181.640625\\
14	185.546875\\
15	191.40625\\
16	193.359375\\
17	199.21875\\
18	208.984375\\
19	218.75\\
20	234.375\\
21	250\\
22	255.859375\\
23	255.859375\\
24	242.1875\\
25	230.46875\\
26	230.46875\\
27	236.328125\\
28	234.375\\
29	236.328125\\
30	240.234375\\
31	240.234375\\
32	232.421875\\
33	214.84375\\
34	203.125\\
35	203.125\\
36	214.84375\\
37	224.609375\\
38	230.46875\\
39	232.421875\\
40	228.515625\\
41	218.75\\
42	216.796875\\
43	208.984375\\
44	205.078125\\
45	203.125\\
46	199.21875\\
47	197.265625\\
48	191.40625\\
49	195.3125\\
50	205.078125\\
51	210.9375\\
52	220.703125\\
53	234.375\\
54	251.953125\\
55	265.625\\
56	285.15625\\
57	287.109375\\
58	273.4375\\
59	257.8125\\
60	230.46875\\
61	185.546875\\
62	181.640625\\
63	185.546875\\
64	183.59375\\
65	173.828125\\
66	173.828125\\
67	175.78125\\
68	173.828125\\
69	169.921875\\
70	167.96875\\
71	164.0625\\
72	160.15625\\
73	160.15625\\
74	158.203125\\
75	156.25\\
76	154.296875\\
77	150.390625\\
78	142.578125\\
79	134.765625\\
80	138.671875\\
81	150.390625\\
82	148.4375\\
83	142.578125\\
84	140.625\\
85	142.578125\\
86	142.578125\\
87	144.53125\\
88	144.53125\\
89	144.53125\\
90	343.75\\
91	142.578125\\
92	140.625\\
93	162.109375\\
94	177.734375\\
95	191.40625\\
96	197.265625\\
97	189.453125\\
98	181.640625\\
99	185.546875\\
100	119.140625\\
101	99.609375\\
102	173.828125\\
};
\end{axis}
\end{tikzpicture}%
 \caption{Fundamental frequency estimates using the proposed method (SNR = $3$ dB). The red line indicates the true fundamental frequency and the blue aterisk denotes the estimated fundamental frequency. } \label{fig:pitch_dual_channel} \end{figure}
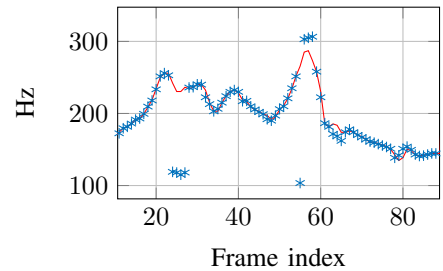

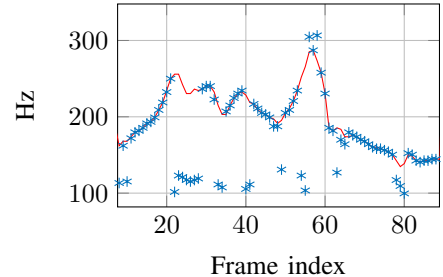
\begin{figure}[t] \centering  \setlength\figureheight{3.0cm} \setlength\figurewidth{4.5cm} 
%
%
\definecolor{mycolor1}{rgb}{0.00000,0.44700,0.74100}%
\begin{tikzpicture}

\begin{axis}[%
width=0.95092\figurewidth,
height=0.6\figurewidth,
at={(0\textwidth,0\textwidth)},
scale only axis,
xmin=7.60368663594469,
xmax=88.8940092165899,
xmajorgrids,
ylabel = Hz,xlabel = Frame index,
ymin=81.9241982507289,
ymax=347.813411078717,
ymajorgrids
]
\addplot [color=mycolor1,only marks,mark=asterisk,mark options={solid},forget plot]
  table[row sep=crcr]{%
1	111.328125\\
2	101.5625\\
3	189.453125\\
4	111.328125\\
5	101.5625\\
6	113.28125\\
7	185.546875\\
8	113.28125\\
9	162.109375\\
10	115.234375\\
11	171.875\\
12	179.6875\\
13	181.640625\\
14	185.546875\\
15	191.40625\\
16	193.359375\\
17	199.21875\\
18	208.984375\\
19	218.75\\
20	232.421875\\
21	250\\
22	101.5625\\
23	123.046875\\
24	121.09375\\
25	117.1875\\
26	115.234375\\
27	117.1875\\
28	119.140625\\
29	236.328125\\
30	240.234375\\
31	240.234375\\
32	222.65625\\
33	111.328125\\
34	107.421875\\
35	207.03125\\
36	214.84375\\
37	224.609375\\
38	230.46875\\
39	234.375\\
40	105.46875\\
41	111.328125\\
42	216.796875\\
43	210.9375\\
44	205.078125\\
45	203.125\\
46	199.21875\\
47	187.5\\
48	187.5\\
49	130.859375\\
50	205.078125\\
51	208.984375\\
52	220.703125\\
53	234.375\\
54	123.046875\\
55	103.515625\\
56	304.6875\\
57	287.109375\\
58	306.640625\\
59	257.8125\\
60	230.46875\\
61	185.546875\\
62	181.640625\\
63	126.953125\\
64	169.921875\\
65	164.0625\\
66	179.6875\\
67	175.78125\\
68	173.828125\\
69	169.921875\\
70	167.96875\\
71	164.0625\\
72	160.15625\\
73	158.203125\\
74	158.203125\\
75	156.25\\
76	154.296875\\
77	150.390625\\
78	117.1875\\
79	109.375\\
80	99.609375\\
81	152.34375\\
82	150.390625\\
83	142.578125\\
84	140.625\\
85	142.578125\\
86	142.578125\\
87	144.53125\\
88	144.53125\\
89	144.53125\\
90	142.578125\\
91	142.578125\\
92	332.03125\\
93	167.96875\\
94	117.1875\\
95	130.859375\\
96	99.609375\\
97	132.8125\\
98	113.28125\\
99	136.71875\\
100	109.375\\
101	125\\
102	99.609375\\
};
\addplot [color=red,solid,forget plot]
  table[row sep=crcr]{%
1	132.8125\\
2	179.6875\\
3	99.609375\\
4	173.828125\\
5	166.015625\\
6	171.875\\
7	187.5\\
8	162.109375\\
9	167.96875\\
10	167.96875\\
11	173.828125\\
12	179.6875\\
13	181.640625\\
14	185.546875\\
15	191.40625\\
16	193.359375\\
17	199.21875\\
18	208.984375\\
19	218.75\\
20	234.375\\
21	250\\
22	255.859375\\
23	255.859375\\
24	242.1875\\
25	230.46875\\
26	230.46875\\
27	236.328125\\
28	234.375\\
29	236.328125\\
30	240.234375\\
31	240.234375\\
32	232.421875\\
33	214.84375\\
34	203.125\\
35	203.125\\
36	214.84375\\
37	224.609375\\
38	230.46875\\
39	232.421875\\
40	228.515625\\
41	218.75\\
42	216.796875\\
43	208.984375\\
44	205.078125\\
45	203.125\\
46	199.21875\\
47	197.265625\\
48	191.40625\\
49	195.3125\\
50	205.078125\\
51	210.9375\\
52	220.703125\\
53	234.375\\
54	251.953125\\
55	265.625\\
56	285.15625\\
57	287.109375\\
58	273.4375\\
59	257.8125\\
60	230.46875\\
61	185.546875\\
62	181.640625\\
63	185.546875\\
64	183.59375\\
65	173.828125\\
66	173.828125\\
67	175.78125\\
68	173.828125\\
69	169.921875\\
70	167.96875\\
71	164.0625\\
72	160.15625\\
73	160.15625\\
74	158.203125\\
75	156.25\\
76	154.296875\\
77	150.390625\\
78	142.578125\\
79	134.765625\\
80	138.671875\\
81	150.390625\\
82	148.4375\\
83	142.578125\\
84	140.625\\
85	142.578125\\
86	142.578125\\
87	144.53125\\
88	144.53125\\
89	144.53125\\
90	343.75\\
91	142.578125\\
92	140.625\\
93	162.109375\\
94	177.734375\\
95	191.40625\\
96	197.265625\\
97	189.453125\\
98	181.640625\\
99	185.546875\\
100	119.140625\\
101	99.609375\\
102	173.828125\\
};
\end{axis}
\end{tikzpicture}%
 \caption{Fundamental frequency estimates using the corresponding single channel method\cite{christensen2009multi} (SNR = $3$ dB).} \label{fig:pitch_single_channel} \end{figure}

As we have seen previously, the formulation of the state transition matrix in (\ref{eq:exc_state_space}) requires the estimation of pitch parameters.
In this paper, we propose a parametric method to estimate the pitch parameters of clean speech present in noise. The babble noise generally encountered in a cocktail party scenario is spectrally coloured. As the pitch estimator proposed here is optimal only for white Gaussian noise signals,
pre-whitening is first performed on the noisy signal to whiten the noise component. Pre-whitening is performed using the estimated noise AR coefficients as
\begin{equation}
\tilde{z}_{l/r}(n) = z_{l/r}(n) +\sum_{i=1}^{Q}\hat{c}_i(f_n)z_{l/r}(n-i).
\end{equation}
The method proposed here operates on signal vectors $\tilde{\mathbf{z}}_{{l/r}_c}(f_n M) \in \mathbb{C}^M$  defined as $\tilde{\mathbf{z}}_{{l/r}_c}(f_n M) = [\tilde{z}_{{l/r}_c}(f_nM), \hdots, \tilde{z}_{{l/r}_c}(f_nM+M-1)]^T$ where $\tilde{z}_{{l/r}_c}(n)$ is the complex signal corresponding to $\tilde{z}_{{l/r}}(n)$, which is obtained using the Hilbert transform.  This method uses the harmonic model to represent the clean speech as a sum of $L$ harmonically related complex sinusoids. 
Using the harmonic model, the noisy signal at the left ear in vector of Gaussian noise $\tilde{\mathbf{w}}_{l_c}(f_nM)$, with covariance matrix, $\mathbf{Q}_l(f_n)$, is represented as
\begin{equation}
\tilde{\mathbf{z}}_{l_c}(f_nM) = \mathbf{V}(f_n)\mathbf{D}_{l}\mathbf{q}(f_n) + \tilde{\mathbf{w}}_{l_c}(f_nM)
\label{eq: left_harm_noisy}
\end{equation}
where  $\mathbf{q}(f_n)$ is a vector of complex amplitudes, $\mathbf{V}(f_n)$ is the Vandermonde  matrix  defined as $\mathbf{V}(f_n)= [\mathbf{v}_1(f_n) \hdots \mathbf{v}_L(f_n)]$, where $[\mathbf{v}_p(f_n)]_m = e^{\imath\omega_0p(f_nM + m-1)}$ with $\omega_0$ being the fundamental frequency and $\mathbf{D}_{l}$ being the directivity matrix from the source to the left ear. The directivity matrix contains a frequency and angle dependent delay and magnitude term along the diagonal,  designed using the method in \cite[eq. 3]{brown1998structural}. Similarly, the noisy signal at the right ear is written as
\begin{equation}
\tilde{\mathbf{z}}_{r_c}(f_nM) = \mathbf{V}(f_n)\mathbf{D}_{r}\mathbf{q}(f_n) + \tilde{\mathbf{w}}_{r_c}(f_nM).
\label{eq: right_harm_noisy}
\end{equation}
The frame index $f_n$ will be omitted for the remainder of the section for notational convenience.
Assuming independence between the channels, the likelihood, due to Gaussianity can be expressed as

\begin{equation}
p(\tilde{\mathbf{z}}_{l_c},\tilde{\mathbf{z}}_{r_c} | \boldsymbol{\epsilon}) 
= \mathcal{CN}(\tilde{\mathbf{z}}_{l_c}; \mathbf{V}\mathbf{D}_l \mathbf{q},\mathbf{Q}_l) \,\mathcal{CN}(\tilde{\mathbf{z}}_{r_c}; \mathbf{V}\mathbf{D}_r \mathbf{q},\mathbf{Q}_r)
\end{equation}
where $\boldsymbol{\epsilon}$ is the parameter set containing  $\omega_0$, the complex amplitudes,  the directivity matrices and the noise covariance matrices.
Assuming that the noise is white in both the channels, the likelihood is rewritten as

\begin{equation}
p(\tilde{\mathbf{z}}_{l_c},\tilde{\mathbf{z}}_{r_c} | \boldsymbol{\epsilon}) =
\frac{e^{-\left(\frac{||\tilde{\mathbf{z}}_{{l}_c} - \mathbf{V}\mathbf{D}_{l}\mathbf{q}||^2}{\sigma_l^2}+\frac{||\tilde{\mathbf{z}}_{{r}_c} - \mathbf{V}\mathbf{D}_{r}\mathbf{q}||^2}{\sigma_r^2}\right)}}{(\pi \sigma_l \sigma_r)^{2M}}
\end{equation}
and the log-likelihood is then
\begin{equation}
\label{eq:loglik_pitch}
\begin{split}
\ln p(\tilde{\mathbf{z}}_{l_c},&\tilde{\mathbf{z}}_{r_c} | \boldsymbol{\epsilon}) = -M(\ln\pi\sigma_l^2 + \ln\pi\sigma_r^2)\\&-\left(\frac{||\tilde{\mathbf{z}}_{{l}_c} - \mathbf{V}\mathbf{D}_{l}\mathbf{q}||^2}{\sigma_l^2} + \frac{||\tilde{\mathbf{z}}_{{r}_c} - \mathbf{V}\mathbf{D}_{r}\mathbf{q}||^2}{\sigma_r^2}\right).
\end{split}
\end{equation}
Assuming the fundamental frequency to be known, the ML estimate of the amplitudes is obtained as
\begin{equation}
\hat{\mathbf{q}} = (\mathbf{H}^H\mathbf{H})^{-1}\mathbf{H}^H\mathbf{y},
\end{equation}
where $\mathbf{H} = \begin{bmatrix}
\left(\mathbf{V}\mathbf{D}_{l}\right)^T \,\, \left(\mathbf{V}\mathbf{D}_{r}\right)^T
\end{bmatrix}^T$ and $\mathbf{y} = [\tilde{\mathbf{z}}_{l_c}^T \tilde{\mathbf{z}}_{r_c}^T]^T$.
These amplitude estimates are further used to estimate the noise variances as 
\begin{equation}
\hat{\sigma}_{l/r}^2 = \frac{1}{M}||\hat{\tilde{\mathbf{w}}}_{{l/r}_c}||^2 = \frac{1}{M}|| \tilde{\mathbf{z}}_{{l/r}_c} - \mathbf{V}\mathbf{D}_{l/r}\hat{\mathbf{q}}||^2 .
\end{equation}
Substituting these into (\ref{eq:loglik_pitch}), we obtain the log-likelihood as
\begin{equation}
\ln p(\tilde{\mathbf{z}}_{l_c},\tilde{\mathbf{z}}_{r_c} | \boldsymbol{\epsilon}) \myeq -M( \ln\hat{\sigma}_{l}^2 + \ln\hat{\sigma}_{r}^2 ).
\end{equation}
The ML estimate of the fundamental frequency is then
\begin{equation}
\label{eq: cost_func}
\hat{\omega}_0 = \argmin_{  \omega_0 \in \Omega_0  } \,\, (\ln\hat{\sigma}_{l}^2 + \ln\hat{\sigma}_{r}^2),
\end{equation}
where $\Omega_0$ is the set of candidate fundamental frequencies. This leads to (\ref{eq: cost_func}) being evaluated on grid of candidate fundamental frequencies. The pitch is then obtained by rounding the reciprocal of the estimated fundamental frequency in Hz. We remark that the model order $L$ is estimated here using the maximum a posteriori (MAP) rule \cite[p. 38]{christensen2009multi}.
 The degree of voicing is calculated by taking the ratio between the energy (calculated as the square of the $l^2$-norm) present at  integer multiples of the fundamental frequency and the total energy present in the signal. This is motivated by the observation that, in case of highly voiced regions, the energy of the signal will be concentrated at the harmonics.
Figures \ref{fig:pitch_dual_channel} and \ref{fig:pitch_single_channel} show the pitch estimation plot from the binaural noisy signal (SNR  = $3$ dB) for the proposed method (which uses information from the two channels), and a single channel pitch estimation method which uses only the left channel, respectively. The red line denotes the true fundamental frequency and the blue asterisk denotes the estimated fundamental frequency. It can be seen that the use of the two channels leads to a more robust pitch estimation.

The main steps involved in the proposed enhancement framework for the V-UV model are shown in Algorithm \ref{alg:Algorithm}. The enhancement framework for the UV model differs from the V-UV model in that it does not require  estimation of the pitch parameters, and that the FLKS equations would be derived based on (\ref{eq: final state space_UV}) and (\ref{eq: final observation_UV}) instead of (\ref{eq: final_state_space_VUV}) and (\ref{eq: final observation_V_UV}).

\begin{algorithm}[t]
	\caption{Main steps involved in the binaural enhancement framework}
	\label{alg:Algorithm}
	\begin{algorithmic} [1]
		\WHILE{new time-frames are available}
		\STATE Estimate the dual channel noise PSD and append the noise codebook with the AR coefficients corresponding to the estimated noise PSD $\hat{P}^{\text{DC}}_w$ (see Section \ref{ssec: Dual noise psd}). 		
        \FOR{$\forall i \in N_s$} 
        \FOR{$\forall j \in N_w$} 
        \STATE compute the ML estimates of excitation noise variances ($\sigma_{d,ij}^{2} \,\, \text{and} \,\, \sigma_{v,ij}^{2}$) using (\ref{eq:exc_var_ML_1}) and (\ref{eq:exc_var_ML_2}).        
        \STATE compute the modelled spectrum $\hat{P}_{z_{ij}}^{\text{ML}}$ using (\ref{eq:modelled_spectrum}).
        \STATE compute the likelihood values $p(\mathbf{z}_l,\mathbf{z}_r|\boldsymbol{\theta}_{ij}^{\text{ML}})$ using (\ref{eq:final weight}).
        \ENDFOR
        \ENDFOR
        \STATE Get the final estimates of STP parameters using (\ref{eq: final_discrete_MMSE}).
        		\STATE Estimate the pitch parameters using the algorithm explained in Section \ref{ssec:            pitch_estimator}.
        \STATE Use the estimated STP parameters and  the pitch parameters in the FLKS equations (see Appendix \ref{ssec:App_pred}) to get the enhanced signal.
		\ENDWHILE
	\end{algorithmic} 
\end{algorithm}

\section{Simulation Results}
\label{sec : Experiments}

In this section, we will present the  experiments that have been carried out to evaluate the proposed enhancement framework. 
\subsection{Implementation details}
\label{ssec: Implementation details}
 The test audio files used for the experiments consisted of speech from the GRID database \cite{cooke2006audio} re-sampled to 8 kHz. The noisy signals were generated using the simulation set-up  explained in Section \ref{ssec : simulation setup}. 
  The speech and noise STP parameters required for the enhancement process were estimated every 25 ms using the codebook-based approach, as explained in Section \ref{ssec : codebook}. The speech codebook and noise codebook used for the estimation of the STP parameters are obtained by the generalised Lloyd algorithm \cite{linde1980algorithm}. During the training process, AR coefficients  (converted into line spectral frequency coefficients) are extracted from windowed frames, obtained from the training signal and passed as an input to the vector quantiser. Working in the line spectral frequency domain is guaranteed to result in stable inverse filters \cite{gray1976distance}. Codebook vectors are then obtained as an output from the vector quantiser depending on the size of the codebook. For our experiments, we have used both a speaker-specific codebook and a general speech codebook. A speaker-specific codebook of 64 entries was generated using head related impulse response (HRIR) convolved speech from the specific speaker of interest. A general speech codebook of 256 entries was generated from a training sample of 30 minutes of HRIR convolved speech from 30 different speakers. Using a speaker-specific codebook  instead of a general speech codebook leads to an improvement in performance, and a comparison between the two was made in \cite{mathew2016}. It should be noted that the sentences used for training the codebook were not included in the test sequence. The noise codebook consisting of only 8 entries, was generated using thirty seconds of noise signal \cite{etsi_binaural}. The AR model order for both the speech and noise signal was empirically chosen to be 14. The pitch period and degree of voicing was estimated as explained in Section \ref{ssec: pitch_estimator} where the cost function in (\ref{eq: cost_func}) was evaluated on a $0.5$ Hz grid for fundamental frequencies in the range $80-400$ Hz. For each fundamental frequency candidate $\omega_0$, the model orders considered were $\mathcal{L} =  \{1,\hdots, \floor*{2\pi / \omega_0}\}$. 
\subsection{Simulation set-up}
\label{ssec : simulation setup}
In this paper we have considered two simulation set-ups representative of the cocktail party scenario. The details regarding the two set-ups are given below:

\subsubsection{Set-up 1}
The clean signals were at first convolved with an anechoic binaural HRIR corresponding to the nose direction, taken from a database \cite{kayser2009database}. Noisy signals are then generated by adding binaurally recorded babble noise taken from the ETSI database \cite{etsi_binaural}.

\subsubsection{Set-up 2}
\begin{figure}
\centering
\includegraphics[width=6cm, height=4cm]{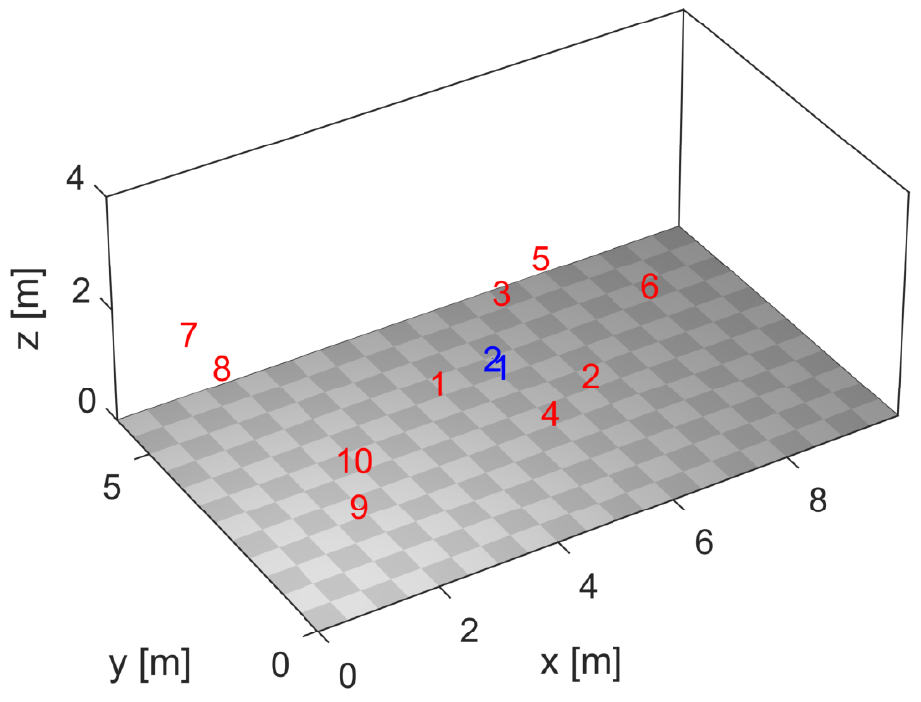}
\caption{Set-up 2 showing the cocktail scenario where 1 (red) indicates the speaker of interest and 2-10 (red) are the interferers and 1,2 (blue) are the microphones on the left ear and right ear respectively. }
\label{fig: setup2}
\end{figure}

The noisy signals were generated using the McRoomSim acoustic simulation software \cite{wabnitz2010room}. Fig. \ref{fig: setup2} shows the geometry of the room along with the speaker, listener and the interferers. This denotes a typical cocktail party scenario, where 1 (red) indicates the speaker of interest, 2-10 (red) are the interferers, and 1, 2 (blue) are the microphones on the left, right ears respectively. The dimensions of the room in this case is $10\times6\times4$ $m$. The reverberation time of the room was chosen to be 0.4 $s$. 
\subsection{Evaluated enhancement frameworks}
In this section we will give an overview about the binaural and bilateral enhancement frameworks that have been evaluated in this paper using the objective and subjective scores.
\subsubsection{Binaural enhancement framework}
In the binaural enhancement framework, we assume that there is a wireless link between the HAs. Thus, the filter parameters are estimated jointly using the information at the left and right channels.
\begin{enumerate}
\item[] \textit{Proposed methods} :  The binaural enhancement framework utilising the V-UV model, when used in conjunction with a general speech codebook is denoted as Bin-S(V-UV), whereas Bin-Spkr(V-UV) denotes the case where we use a speaker-specific codebook.
 The binaural enhancement framework utilising the UV model, when used in conjunction with a general speech codebook is denoted as Bin-S(UV), whereas Bin-Spkr(UV) denotes the case where we use a speaker-specific codebook.
\vspace{1.2mm}
\item[] \textit{Reference methods} : For comparison, we have used the methods proposed in  \cite{dorbecker1996combination} and \cite{li2011two} which we denote as  TwoChSS and TS-WF  respectively. We chose  these methods for comparison, as TwoChSS was one of the first methods designed for a two-input two-output configuration and TS-WF is one of the state of the art methods belonging to this class.

\end{enumerate}

\subsubsection{Bilateral enhancement framework}
In the bilateral enhancement framework, single channel speech enhancement techniques are performed independently on each ear.
\begin{enumerate}
\item[] \textit{Proposed methods} : The bilateral enhancement framework utilising the V-UV model, when used in conjunction with a general speech codebook is denoted as Bil-S(V-UV), whereas Bil-Spkr(V-UV) denotes the case where we use a speaker-specific codebook.
 The bilateral enhancement framework utilising the UV model, when used in conjunction with a general speech codebook is denoted as Bil-S(UV), whereas Bil-Spkr(UV) denotes the case where we use a speaker-specific codebook. The difference of the bilateral case in comparison to the binaural case is in the estimation of the filter parameters. 
 In the bilateral case, the filter parameters are estimated independently for each ear which leads to different filter parameters for each ear, e.g., the STP parameters are estimated using the method in \cite{he2017multiplicative} independently for each ear.
\vspace{1.2mm}
\item[] \textit{Reference methods} : For comparison, we have used the methods proposed in \cite{erkelens2007minimum} and \cite{loizou2005speech} which we denote as MMSE-GGP and PMBE respectively.

\end{enumerate}

\subsection{Objective measures}
\label{ssec: objective_measure}

The objective measures, STOI \cite{taal2011algorithm} and PESQ \cite{recommendation2001perceptual} have been used to evaluate the intelligibility and quality of different enhancement frameworks. We have evaluated the performance of the algorithms, separately for the 2 different simulation set-ups explained in Section \ref{ssec : simulation setup}. Table \ref{table:Bin_setup1 } and \ref{table:Bil_setup1 } show the objective measures obtained for the binaural and bilateral enhancement frameworks, respectively, when evaluated in the  set-up 1. The test signals that have been used for the binaural and bilateral enhancement frameworks are identical. The scores shown in the tables are the averaged scores across the left and right channels. In comparison to the reference methods which reduce the STOI scores, it can be seen that all of the proposed methods improve the STOI scores. It can be seen from Tables \ref{table:Bin_setup1 } and \ref{table:Bil_setup1 } that the Bin-Spkr(V-UV) performs the best in terms of STOI scores. In addition to preserving the binaural cues, it is evident from the scores that the binaural frameworks perform in general better than the bilateral frameworks, and the improvement of binaural framework over bilateral framework is more pronounced at low SNRs. It can also be seen that the V-UV model which takes into account the pitch information performs better than the UV model.  Tables \ref{table:Bin_setup2 } and \ref{table:Bil_setup2 } show the objective measures obtained for the different binaural and bilateral enhancement frameworks, respectively, when evaluated in the simulation set-up 2. The results obtained for set-up 2 shows similar trends to the results obtained for set-up 1. We would also like to remark here that in the range of 0.6-0.8, an increase in $0.05$ in STOI score corresponds to approximately  $16$ percentage points increase in subjective intelligibility \cite{falk2015objective}.

\subsection{Inter-aural errors}
\begin{figure}[h!]
\centering
\begin{subfigure}{.2475\textwidth}
  \centering
  \includegraphics[width=4.5cm, height=3.5cm]{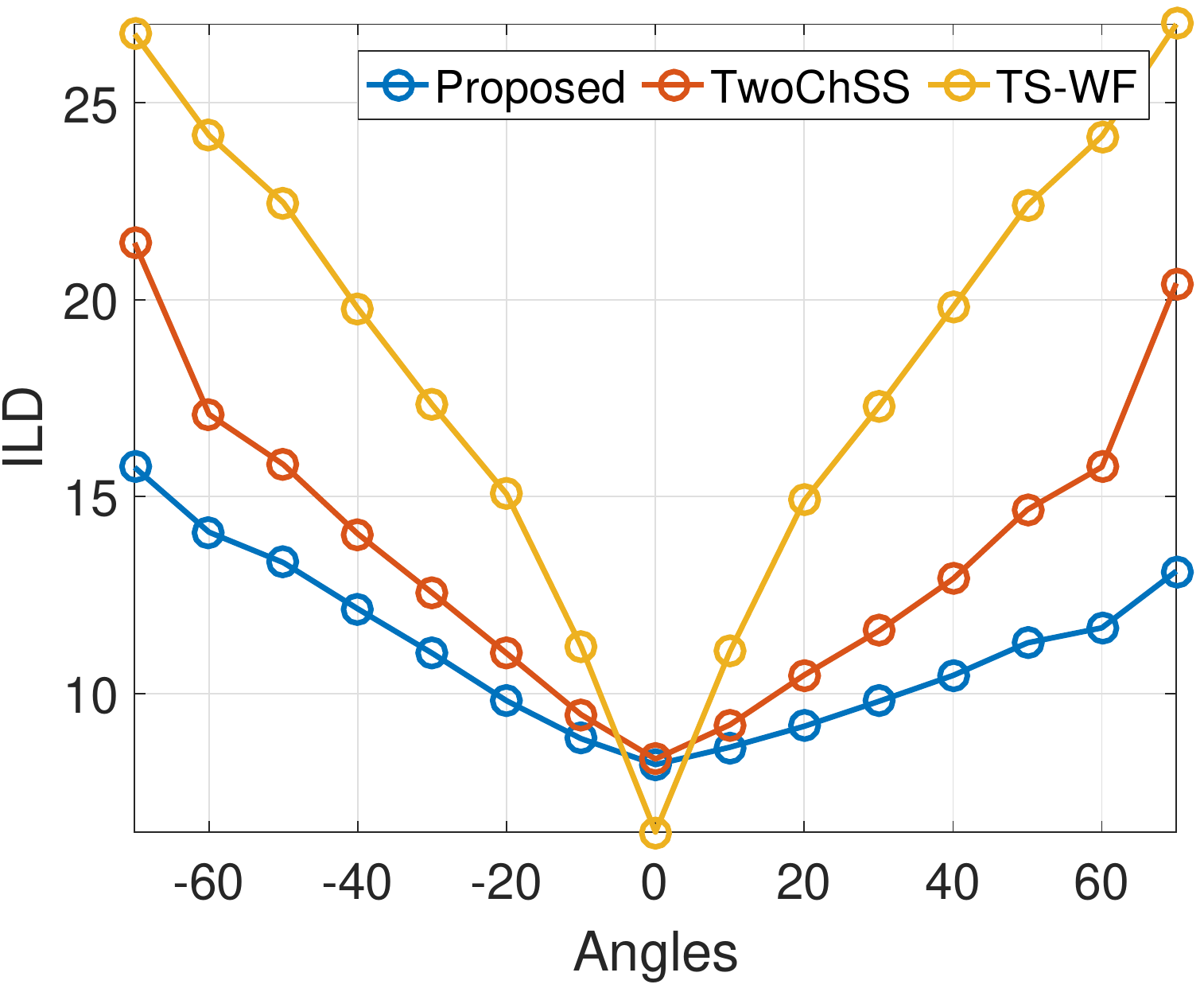}
  \caption{ILD}
  \label{fig:sub1}
\end{subfigure}%
\begin{subfigure}{.2475\textwidth}
  \centering
  \includegraphics[width=4.2cm, height=3.5cm]{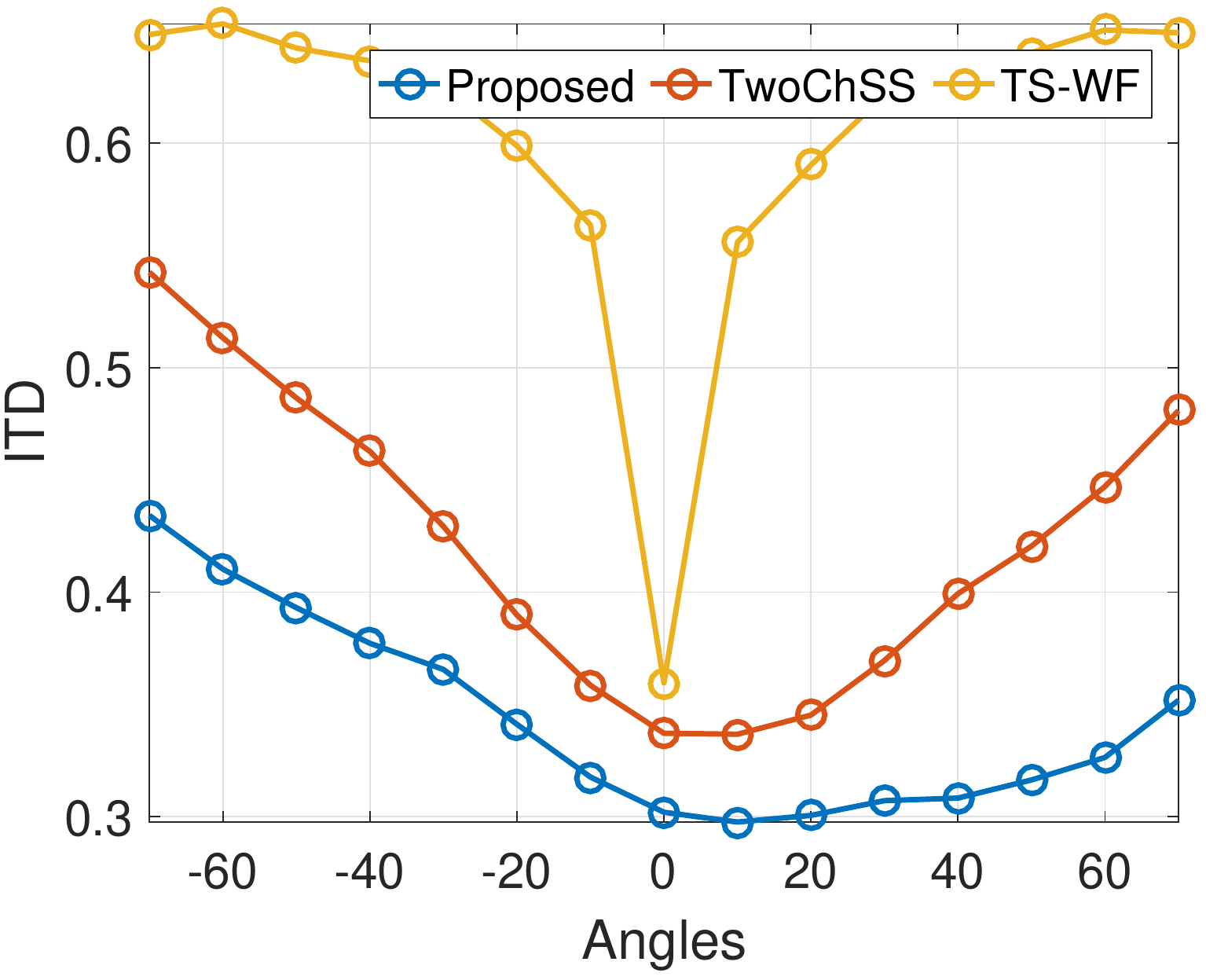}
  \caption{ITD}
  \label{fig:sub2}
\end{subfigure}
\caption{Inter-aural cues for different speaker positions.}
\label{fig: inter_au}
\end{figure}

\!
We now evaluate the proposed algorithm in terms of binaural cue preservation. This was evaluated objectively using inter-aural time difference (ITD) and inter-aural level difference (ILD) also used in \cite{li2011two}. ITD is calculated as
\begin{equation}
\text{ITD} = \frac{|\angle C_{\text{enh}} - \angle C_{\text{clean}}|}{\pi},
\end{equation}
where $\angle C_{\text{enh}}$ and $\angle C_{\text{clean}}$ denotes the phases of the cross PSD of the enhanced and clean signal respectively, given by
$
C_{\text{enh}} = \mathbb{E}\{\hat{S}_l\hat{S}_r\}   \,\,\,\text{and} \,\,\, C_{\text{clean}} = \mathbb{E}\{S_lS_r\},
$
where $\hat{S}_{l/r}$ denotes the spectrum of enhanced signal at the left/right ear and $S_{l/r}$ denotes the spectrum of the clean signal at the left/right ear. The expectation is calculated by taking the average value over all frames and frequency indices (which has been omitted here for notational convenience). ILD is calculated as
\begin{equation}
\text{ILD} = \left|10\text{log}_\text{10} \frac{I_{\text{enh}}}{I_{\text{clean}}} \right|,
\end{equation}
where 
$
I_{\text{enh}} = \frac{\mathbb{E}\{|\hat{S}_l|^2\}}{\mathbb{E}\{|\hat{S}_r|^2\}} \,\,\,\text{and} \,\,\, I_{\text{clean}} = \frac{\mathbb{E}\{|S_l|^2\}}{\mathbb{E}\{|S_r|^2\}}.
$
Fig. \ref{fig: inter_au} shows the ILD and ITD cues for the proposed method, Bin-Spkr(V-UV), TwoChSS and TS-WF for different angles of arrivals. It can be seen that the proposed method has a lower ITD and ILD in comparison to TwoChSS and TS-WF. It should be noted that the proposed method and TwoChSS do not use the angle of arrival and assume that the speaker of interest is in the nose direction of the listener. TS-WF, on the other hand requires the a priori knowledge of the angle of arrival. Thus, to make a fair comparison we have included here the inter-aural cues for TS-WF when the speaker of interest is assumed to be in the nose direction.

\begin{table*}[t]
  \centering
  \caption{This table shows the comparison of objective measures (PESQ \& STOI) for the different \textbf{BINAURAL} enhancement frameworks for 4 different signal to noise ratios. Noisy signals used for the evaluation here is generated using the simulation set-up 1.}
\begin{tabular}{@{}l c c c c c c c  c @{}} \toprule[1.1pt]
                & &  Bin-Spkr(UV) & Bin-Spkr(V-UV) & Bin-S(UV)  &  Bin-S(V-UV)  & TS-WF  &   TwoChSS     &  Noisy   \\ \midrule
                STOI & 0 dB & $0.71$& $\mathbf{0.75}$ &$0.68$  & $0.72$ &  $0.62$& $0.64$ &    $0.67$  \\ 
                 & 3 dB&  $0.80$ & $\mathbf{0.82}$ & $0.77$ & $0.79$ & $0.69$ & $0.72$    &  $0.73$  \\ 
                 & 5 dB&  $0.84$ & $\mathbf{0.85}$& $0.81$ & $0.83$ & $0.74$& $0.77$    &  $0.78$  \\ 
                 & 10 dB&  $0.91$ & $\mathbf{0.91}$ & $0.90$ & $0.90$ & $0.85$ & $0.86$ &   $0.87$  \\ \midrule
 PESQ & 0 dB & $1.43$ & $\mathbf{1.53}$ & $1.37$  & $1.45$ & $1.40$ & $1.49$ &     $1.33$  \\ 
 & 3 dB& $1.67$  & $\mathbf{1.72}$ & $1.58$ &$1.68$ & $1.55$ & $1.66$ &  $1.43$  \\ 
 & 5 dB& $1.80$  & $\mathbf{1.85}$ & $1.73$ & $1.78$ & $1.68$ & $1.79$ &  $1.50$  \\ 
 & 10dB & $\mathbf{2.24}$  & $2.22$ & $2.13$ &$2.14$ & $2.13$ & $2.20$ &  $1.70$  \\ 
                 \bottomrule[1.1pt]
\end{tabular}
\label{table:Bin_setup1 } 
\end{table*}

\begin{table*}[t]
  \centering
  \caption{This table shows the comparison of objective measures (PESQ \& STOI) for the different \textbf{BILATERAL} enhancement frameworks for 4 different signal to noise ratios. Noisy signals used for the evaluation here is generated using the simulation set-up 1.}
\begin{tabular}{@{}l c c c c c c c c  @{}} \toprule[1.1pt]
                & &  Bil-Spkr(UV) & Bil-Spkr(V-UV) & Bil-S(UV)  &  Bil-S(V-UV)  & MMSE-GGP  &  PMBE  &   Noisy   \\ \midrule
 STOI & 0 dB & $0.68$ & $\mathbf{0.72}$ & $0.66$  & $0.70$ & $0.66$ & $0.66$ &    $0.67$  \\ 
 & 3 dB&  $0.77$ & $\mathbf{0.79}$ & $0.75$ & $0.78$ & $0.73$ & $0.73$ & $0.73$  \\ 
 & 5 dB&  $0.81$ & $\mathbf{0.83}$ & $0.80$ & $0.82$ & $0.78$ & $0.78$ &    $0.78$  \\ 
 & 10 dB&  $0.90$ & $\mathbf{0.90}$ & $0.89$ & $0.90$ & $0.87$ & $0.87$&     $0.87$  \\ \midrule
  PESQ & 0 dB & $1.37$ & $\mathbf{1.45}$ & $1.34$  & $1.40$ & $1.26$ & $1.30$ &     $1.33$  \\ 
  & 3 dB& $1.58$  & $\mathbf{1.65}$ & $1.53$ &$1.60$ & $1.43$ & $1.43$ &     $1.43$  \\ 
  & 5 dB& $1.72$  & $\mathbf{1.76}$ & $1.66$ & $1.72$ & $1.50$ & $1.56$ &    $1.50$  \\ 
  & 10 dB & $\mathbf{2.12}$  & $2.10$ & $2.04$ & $2.05$ & $1.73$ & $1.79$ &    $1.70$  \\ 
      
                 \bottomrule[1.1pt]
\end{tabular}
 \label{table:Bil_setup1 } 
\end{table*}

\begin{table*}[t]
  \centering
  \caption{This table shows the comparison of STOI scores for the different \textbf{BINAURAL} enhancement frameworks for 4 different signal to noise ratios. Noisy signals used for the evaluation here is generated using the simulation set-up 2.}
\begin{tabular}{@{}l c c c c c c c  c @{}} \toprule[1.1pt]
                & &  Bin-Spkr(UV) & Bin-Spkr(V-UV) & Bin-S(UV)  &  Bin-S(V-UV)  & TS-WF  &   TwoChSS     &  Noisy   \\ \midrule
                STOI & 0 dB & $0.63$& $\mathbf{0.68}$ &$0.61$  & $0.66$ &  $0.62$& $0.58$ &    $0.60$  \\ 
                 & 3 dB&  $0.73$ & $\mathbf{0.75}$ & $0.71$ & $0.74$ & $0.69$ & $0.67$    &  $0.68$  \\ 
                 & 5 dB&  $0.78$ & $\mathbf{0.80}$& $0.76$ & $0.79$ & $0.73$& $0.72$    &  $0.73$  \\ 
                 & 10 dB&  $0.88$ & $\mathbf{0.89}$ & $0.87$ & $0.88$ & $0.81$ & $0.83$ &   $0.84$  \\ 
                 \bottomrule[1.1pt]
\end{tabular}
 \label{table:Bin_setup2 } 
\end{table*}

\begin{table*}[t]
  \centering
  \caption{This table shows the comparison of STOI scores for the different \textbf{BILATERAL} enhancement frameworks for 4 different signal to noise ratios. Noisy signals used for the evaluation here is generated using the simulation set-up 2.}
\begin{tabular}{@{}l c c c c c c c c  @{}} \toprule[1.1pt]
                & &  Bil-Spkr(UV) & Bil-Spkr(V-UV) & Bil-S(UV)  &  Bil-S(V-UV)  & MMSE-GGP  &  PMBE  &   Noisy   \\ \midrule
 STOI & 0 dB & $0.61$ & $\mathbf{0.65}$ & $0.60$  & $0.64$ & $0.58$ & $0.60$ &    $0.60$  \\ 
 & 3 dB&  $0.71$ & $\mathbf{0.74}$ & $0.69$ & $0.73$ & $0.66$ & $0.68$ & $0.68$  \\ 
 & 5 dB&  $0.76$ & $\mathbf{0.79}$ & $0.75$ & $0.78$ & $0.72$ & $0.73$ &    $0.73$  \\ 
 & 10 dB&  $0.87$ & $\mathbf{0.88}$ & $0.86$ & $0.88$ & $0.83$ & $0.84$ &     $0.84$  \\ 
  \bottomrule[1.1pt]
\end{tabular}
 \label{table:Bil_setup2 } 
\end{table*}

\subsection{Listening tests}
\label{ssec: subjective_tests}
We have conducted listening tests to measure the performance of the proposed algorithm in terms of quality and intelligibility improvements. The tests were conducted on a set of nine NH subjects. These tests were performed in a silent room using a set of Beyerdynamic DT 990 pro headphones. The speech enhancement method that we have evaluated in the listening tests is Bil-Spkr(V-UV) for a single channel. We chose this case for the tests as  we wanted to test the simpler, but more challenging case of intelligibility and quality improvement when we have access to only a single channel. Moreover, as the tests were conducted with NH subjects, we also wanted to eliminate any bias in the results that can be caused due to the binaural cues \cite{bronkhorst1990clinical}, as the benefit of using binaural cues is higher for a NH person than for a hearing impaired person.

\subsubsection{Quality tests}
\begin{figure}
\centering
\includegraphics[width=7.5cm, height=3.8cm]{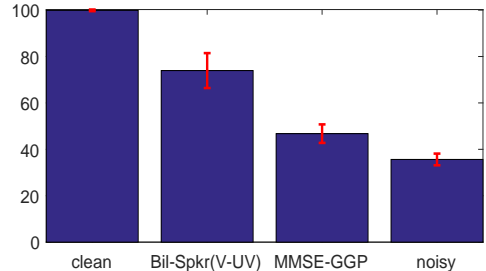}
\caption{Figure showing the mean scores and the $95\%$ confidence intervals obtained in the MUSHRA test for the different methods.}
\label{fig: mushra}
\end{figure}

Quality performance of the proposed algorithms were evaluated using MUSHRA experiments \cite{recommendation20031534}. The test subjects  were asked to evaluate the quality of the processed audio-files using a MUSHRA set-up. The subjects were presented with the clean, processed and the noisy signals. The processing algorithms considered here are Bil-Spkr(V-UV) and MMSE-GGP. The SNR of the noisy signal considered here was $10$ dB. The subjects were then asked to rate the presented signals in a score range of $0-100$. Fig. \ref{fig: mushra} shows the mean scores along with $95\%$ confidence intervals that were obtained for the different methods. It can be seen from the figure that the proposed method performs significantly better than the reference method.
\subsubsection{Intelligibility tests}
\label{ssec: intelligibilty tests}
Intelligibility tests were conducted using sentences from the GRID database \cite{cooke2006audio}. The GRID database contains sentences spoken by 34 different speakers (18 males and 16 females). The sentences are of the following syntax: Bin Blue (Color) by S (Letter) 5 (Digit) please. Table \ref{table:sentence_syntax} shows the syntax of all the possible sentences. subjects are  asked to  identify the color, letter and number after listening to the sentence. The sentences are played back in the SNR range $-8$ to $0$ dB for different algorithms. This SNR range is chosen as all the subjects were NH which led to the intelligibility of the unprocessed signal above $2$ dB to be close to $100$\%. A total of nine test subjects were used for the experiments and the average time taken for carrying out the listening test for a particular person was approximately two hours.  The noise signal that we have used for the tests is the babble signal from the AURORA database \cite{hirsch2000aurora}. The test subjects evaluated the noisy signals ($unp$) and two versions of the processed signal, $nr_{100}$ and $nr_{85}$. The first version, $nr_{100}$, refers to the completely enhanced signal and the second version, $nr_{85}$, refers to a mixture of the enhanced signal and the noisy signal with $85\%$ of the enhanced signal and $15\%$ of the noisy signal. This mixing combination was empirically chosen \cite{anzalone2006determination}. Figures \ref{fig: digit_acc}, \ref{fig: color_acc} and \ref{fig: letter_acc} show the intelligibility percentage along with $90\%$ probability intervals obtained for digit, color and the letter field respectively as a function of SNR, for the different methods. It can be seen that $nr_{85}$ performs the best consistently followed by $nr_{100}$ and the $unp$. Fig. \ref{fig: mean_acc} shows the mean accuracy over all the 3 fields. It can be seen from the figure that $nr_{85}$ gives up to $15\%$ improvement in intelligibility at $-8$ dB SNR. We have also computed the probabilities that a particular method is better than the unprocessed signal in terms of intelligibility. For the computation of these probabilities, the posterior probability of success for each method is modelled using a beta distribution. Table \ref{table : Intelligibilty} shows these probabilities at different SNRs for the $3$ different fields. $P(nr_{85}>unp)$ denotes the probability that $nr_{85}$ is better than $unp$. It can be seen from the table that $nr_{85}$ consistently has a very high probability of being better than $unp$ for all the SNRs, whereas $nr_{100}$ has a high probability of decreasing the intelligibility for the color field at $-2$ dB and the letter field at  $0$ dB. This can also be seen from Figures \ref{fig: color_acc} and \ref{fig: letter_acc}. In terms of the mean intelligibility across all fields, it can be seen that the probability that $nr_{85}$ performs better than $unp$ is $1$ for all the SNRs. Similarly, the probability that $nr_{100}$ also performs better than $unp$ is very high across all SNRs. 

\begin{table}
\caption{Sentence syntax of the GRID database.}
\label{table:sentence_syntax}
\begin{tabular}{ |p{1.0cm}|p{1.0cm}|p{1.0cm}|p{1.0cm}|p{1.0cm} | p{1.0cm}|  }
 \hline
 \multicolumn{6}{|c|}{Sentence structure} \\
 \hline
\scriptsize{command} & color & \scriptsize{preposition} &letter & digit & adverb\\
 \hline
bin  & blue    &at &  A-Z &0-9 &again\\
 lay&   green  & by   &(no W) & &now\\
 place &red & in &  & &please\\
 set    &white & with& & &soon\\

 \hline
\end{tabular}
\end{table}
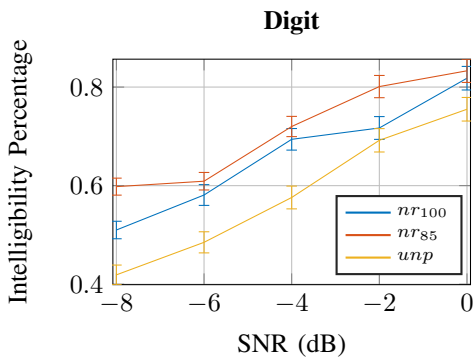
\begin{figure} \centering  \setlength\figureheight{3.0cm} \setlength\figurewidth{5cm} 
%
%
\definecolor{mycolor1}{rgb}{0.00000,0.44700,0.74100}%
\definecolor{mycolor2}{rgb}{0.85000,0.32500,0.09800}%
\definecolor{mycolor3}{rgb}{0.92900,0.69400,0.12500}%
\begin{tikzpicture}

\begin{axis}[%
width=0.95092\figurewidth,
height=\figureheight,
at={(0.758333in,0.48125in)},
scale only axis,
xmin=-8.08,
xmax=0.08,
xlabel={SNR (dB)},
xmajorgrids,
ymin=0.399113700955012,
ymax=0.856592080113072,
ylabel={Intelligibility Percentage},
ymajorgrids,
title style={font=\bfseries},
title={Digit},
legend style={at={(0.622619,0.042063)},anchor=south west,legend cell align=left,align=left,draw=white!15!black}
]
\addplot [color=mycolor1,solid]
 plot [error bars/.cd, y dir = both, y explicit]
 table[row sep=crcr, y error plus index=2, y error minus index=3]{%
-8	0.51	0.0177435267094406	0.0177435267094406\\
-6	0.581	0.0210728553135341	0.0210728553135341\\
-4	0.694	0.0218935389302939	0.0218935389302939\\
-2	0.717	0.0231150694085546	0.0231150694085546\\
0	0.818	0.023901960109581	0.023901960109581\\
};
\addlegendentry{$nr_{100}$};

\addplot [color=mycolor2,solid]
 plot [error bars/.cd, y dir = both, y explicit]
 table[row sep=crcr, y error plus index=2, y error minus index=3]{%
-8	0.598	0.0172368649768595	0.0172368649768595\\
-6	0.609	0.0177748887654953	0.0177748887654953\\
-4	0.72	0.0205331636072184	0.0205331636072184\\
-2	0.801	0.022564829868533	0.022564829868533\\
0	0.833	0.023592080113072	0.023592080113072\\
};
\addlegendentry{$nr_{85}$};

\addplot [color=mycolor3,solid]
 plot [error bars/.cd, y dir = both, y explicit]
 table[row sep=crcr, y error plus index=2, y error minus index=3]{%
-8	0.419	0.0198862990449884	0.0198862990449884\\
-6	0.485	0.0214252202802764	0.0214252202802764\\
-4	0.576	0.0231159348118383	0.0231159348118383\\
-2	0.692	0.0237099093687124	0.0237099093687124\\
0	0.755	0.0239247632971476	0.0239247632971476\\
};
\addlegendentry{$unp$};

\end{axis}
\end{tikzpicture}%
 \caption{Mean percentage of correct answers given by participants for the digit field as function of SNR for different methods. ($unp$) refers to the noisy signal, ($nr_{100}$) refers to the completely enhanced signal and ($nr_{85}$) refers to a mixture of the enhanced signal and the noisy signal with $85\%$ of the enhanced signal and $15\%$ of the noisy signal.} \label{fig: digit_acc} \end{figure}
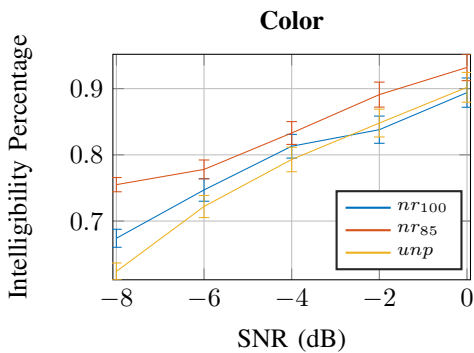
\begin{figure} \centering  \setlength\figureheight{3.0cm} \setlength\figurewidth{5cm} 
%
%
\definecolor{mycolor1}{rgb}{0.00000,0.44700,0.74100}%
\definecolor{mycolor2}{rgb}{0.85000,0.32500,0.09800}%
\definecolor{mycolor3}{rgb}{0.92900,0.69400,0.12500}%
\begin{tikzpicture}

\begin{axis}[%
width=0.95092\figurewidth,
height=\figureheight,
at={(0.758333in,0.48125in)},
scale only axis,
xmin=-8.08,
xmax=0.08,
xlabel={SNR (dB)},
xmajorgrids,
ymin=0.611307088617839,
ymax=0.951886299044988,
ylabel={Intelligibility Percentage},
ymajorgrids,
title style={font=\bfseries},
title={Color},
legend style={at={(0.622619,0.042063)},anchor=south west,legend cell align=left,align=left,draw=white!15!black}
]
\addplot [color=mycolor1,solid]
 plot [error bars/.cd, y dir = both, y explicit]
 table[row sep=crcr, y error plus index=2, y error minus index=3]{%
-8	0.674	0.0136089243453226	0.0136089243453226\\
-6	0.747	0.0170565638473402	0.0170565638473402\\
-4	0.813	0.0179016325231393	0.0179016325231393\\
-2	0.838	0.0205397718836742	0.0205397718836742\\
0	0.894	0.0221257217391834	0.0221257217391834\\
};
\addlegendentry{$nr_{100}$};

\addplot [color=mycolor2,solid]
 plot [error bars/.cd, y dir = both, y explicit]
 table[row sep=crcr, y error plus index=2, y error minus index=3]{%
-8	0.755	0.0107399106260109	0.0107399106260109\\
-6	0.778	0.0142423225586417	0.0142423225586417\\
-4	0.833	0.0172368649768595	0.0172368649768595\\
-2	0.891	0.018873342013662	0.018873342013662\\
0	0.932	0.0198862990449884	0.0198862990449884\\
};
\addlegendentry{$nr_{85}$};

\addplot [color=mycolor3,solid]
 plot [error bars/.cd, y dir = both, y explicit]
 table[row sep=crcr, y error plus index=2, y error minus index=3]{%
-8	0.624	0.012692911382161	0.012692911382161\\
-6	0.722	0.0166775437012171	0.0166775437012171\\
-4	0.793	0.0184844767738798	0.0184844767738798\\
-2	0.848	0.0209926361616302	0.0209926361616302\\
0	0.902	0.022504351085833	0.022504351085833\\
};
\addlegendentry{$unp$};

\end{axis}
\end{tikzpicture}%
 \caption{Mean percentage of correct answers given by participants for color field as function of SNR for different methods.} \label{fig: color_acc} \end{figure}
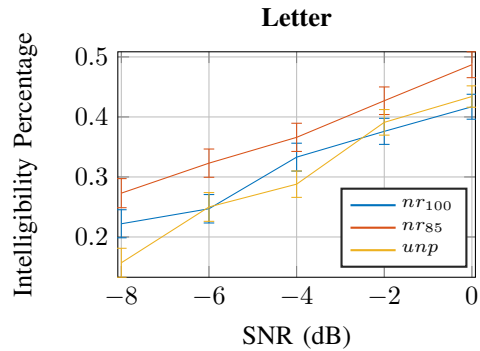
\begin{figure} \centering  \setlength\figureheight{3.0cm} \setlength\figurewidth{5cm} 
%
%
\definecolor{mycolor1}{rgb}{0.00000,0.44700,0.74100}%
\definecolor{mycolor2}{rgb}{0.85000,0.32500,0.09800}%
\definecolor{mycolor3}{rgb}{0.92900,0.69400,0.12500}%
\begin{tikzpicture}

\begin{axis}[%
width=0.95092\figurewidth,
height=\figureheight,
at={(0.758333in,0.48125in)},
scale only axis,
xmin=-8.08,
xmax=0.08,
xlabel={SNR (dB)},
xmajorgrids,
ymin=0.132856980232438,
ymax=0.508561291689954,
ylabel={Intelligibility Percentage},
ymajorgrids,
title style={font=\bfseries},
title={Letter},
legend style={at={(0.622619,0.042063)},anchor=south west,legend cell align=left,align=left,draw=white!15!black}
]
\addplot [color=mycolor1,solid]
 plot [error bars/.cd, y dir = both, y explicit]
 table[row sep=crcr, y error plus index=2, y error minus index=3]{%
-8	0.222	0.0233861879871399	0.0233861879871399\\
-6	0.247	0.0236816932042593	0.0236816932042593\\
-4	0.333	0.0232455066766583	0.0232455066766583\\
-2	0.376	0.021718661130381	0.021718661130381\\
0	0.417	0.0207816243678395	0.0207816243678395\\
};
\addlegendentry{$nr_{100}$};

\addplot [color=mycolor2,solid]
 plot [error bars/.cd, y dir = both, y explicit]
 table[row sep=crcr, y error plus index=2, y error minus index=3]{%
-8	0.273	0.0242318441209707	0.0242318441209707\\
-6	0.323	0.0235367102853954	0.0235367102853954\\
-4	0.366	0.0234792876453866	0.0234792876453866\\
-2	0.427	0.0229955215916477	0.0229955215916477\\
0	0.487	0.0215612916899541	0.0215612916899541\\
};
\addlegendentry{$nr_{85}$};

\addplot [color=mycolor3,solid]
 plot [error bars/.cd, y dir = both, y explicit]
 table[row sep=crcr, y error plus index=2, y error minus index=3]{%
-8	0.157	0.0241430197675618	0.0241430197675618\\
-6	0.25	0.0239653551773951	0.0239653551773951\\
-4	0.288	0.0220251907803479	0.0220251907803479\\
-2	0.391	0.0213070137011127	0.0213070137011127\\
0	0.434	0.0178112986136367	0.0178112986136367\\
};
\addlegendentry{$unp$};

\end{axis}
\end{tikzpicture}%
 \caption{Mean percentage of correct answers given by participants for letter field as function of SNR for different methods.} \label{fig: letter_acc} \end{figure}
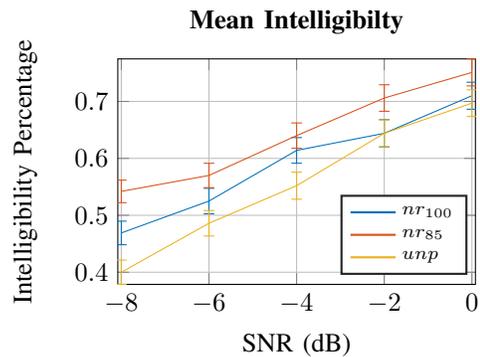
\begin{figure} \centering  \setlength\figureheight{3.0cm} \setlength\figurewidth{5cm} 
%
%
\definecolor{mycolor1}{rgb}{0.00000,0.44700,0.74100}%
\definecolor{mycolor2}{rgb}{0.85000,0.32500,0.09800}%
\definecolor{mycolor3}{rgb}{0.92900,0.69400,0.12500}%
\begin{tikzpicture}

\begin{axis}[%
width=0.95092\figurewidth,
height=\figureheight,
at={(0.758333in,0.48125in)},
scale only axis,
xmin=-8.08,
xmax=0.08,
xlabel={SNR (dB)},
xmajorgrids,
ymin=0.378638917322351,
ymax=0.774724694545439,
ylabel={Intelligibility Percentage},
ymajorgrids,
title style={font=\bfseries},
title={Mean Intelligibilty },
legend style={at={(0.622619,0.042063)},anchor=south west,legend cell align=left,align=left,draw=white!15!black}
]
\addplot [color=mycolor1,solid]
 plot [error bars/.cd, y dir = both, y explicit]
 table[row sep=crcr, y error plus index=2, y error minus index=3]{%
-8	0.469	0.0206869830356209	0.0206869830356209\\
-6	0.525	0.0223862733128083	0.0223862733128083\\
-4	0.614	0.022547302556622	0.022547302556622\\
-2	0.644	0.0239880656543844	0.0239880656543844\\
0	0.71	0.0237235410829011	0.0237235410829011\\
};
\addlegendentry{$nr_{100}$};

\addplot [color=mycolor2,solid]
 plot [error bars/.cd, y dir = both, y explicit]
 table[row sep=crcr, y error plus index=2, y error minus index=3]{%
-8	0.542	0.0198059357029425	0.0198059357029425\\
-6	0.57	0.0214020241489227	0.0214020241489227\\
-4	0.64	0.0222477956912679	0.0222477956912679\\
-2	0.706	0.0232804380119653	0.0232804380119653\\
0	0.751	0.0237246945454388	0.0237246945454388\\
};
\addlegendentry{$nr_{85}$};

\addplot [color=mycolor3,solid]
 plot [error bars/.cd, y dir = both, y explicit]
 table[row sep=crcr, y error plus index=2, y error minus index=3]{%
-8	0.4	0.0213610826776492	0.0213610826776492\\
-6	0.486	0.0223862733128083	0.0223862733128083\\
-4	0.552	0.0237536100216093	0.0237536100216093\\
-2	0.644	0.0235506217381483	0.0235506217381483\\
0	0.697	0.0234461059631359	0.0234461059631359\\
};
\addlegendentry{$unp$};

\end{axis}
\end{tikzpicture}%
 \caption{Mean percentage of correct answers given by participants for all the fields as function of SNR for different methods.} \label{fig: mean_acc} \end{figure}

\section{Discussion}
\label{sec:disc}
The noise reduction capabilities of a HA are limited especially in situations such as the cocktail party scenario. Single channel speech enhancement algorithms which do not use any prior information regarding the speech and noise type have not been able to show much improvements in speech intelligibility \cite{loizou2011reasons}. 
 A class of algorithms that has received significant attention recently have been the deep neural network (DNN) based speech enhancement systems. These algorithms use a priori information about speech and noise types to learn the structure of the mapping function between noisy and clean speech features. These methods were able to show improvements in speech intelligibility when trained to very specific scenarios. Recently, the performance of a general DNN based enhancement system was investigated in terms of objective measures and intelligibility tests \cite{kolbaek2017speech}. Even though the general system  showed improvements in the objective measures, the intelligibility tests failed to show consistent improvements across the SNR range. In this paper we have proposed a model-based speech enhancement framework that takes into account the speech production model, characterised by the vocal tract and the excitation signal. The proposed framework uses a priori information regarding the speech spectral envelopes (which is used for modelling the characteristics of the vocal tract) and noise spectral envelopes. In comparison to DNN based algorithms the training data required by the proposed algorithm, and the parameters to be trained for the proposed algorithm is significantly less. The parameters to be trained in the proposed algorithm includes the AR coefficients corresponding to the speech and noise spectral shapes which is considerably less compared to the weights present in a DNN. As the amount of parameters to be trained is much smaller, it should also be possible to train these parameters on-line in case of noise only scenarios or speech only scenarios. The proposed framework was able to show consistent improvements in the intelligibility tests even for the single channel case as shown in section \ref{ssec: intelligibilty tests}. Moreover, we have shown the benefit of using multiple channels for enhancement by the means of objective experiments.
We would like to remark that the enhancement algorithm proposed in this paper is computationally more complex when compared to conventional speech enhancement algorithms such as \cite{erkelens2007minimum}. However, there exists some methods in the literature which can reduce the computational complexity of the proposed algorithm. The pitch estimation algorithm can be sped up using the principles proposed in \cite{nielsen2017fast}. There also exists efficient ways of performing Kalman filtering due to the structured and sparse matrices involved in the operation of a Kalman filter \cite{goh1999kalman}.
\begin{table}[t]
\centering
\caption{This table shows the probabilities that a particular method is better than the unprocessed signal.}
\begin{tabular}{cc|c|c|c|c|c|l}
\cline{3-7}
& & \multicolumn{5}{ c| }{SNR (dB)} \\ \cline{3-7}
& & -8 & -6 & -4 & -2& 0\\ \cline{1-7}
\multicolumn{1}{ |c  }{\multirow{2}{*}{Digit} } &
\multicolumn{1}{ |c| }{$P(nr_{85}>unp)$} & $1$ & $1$ & $1$ & $1$ &$ 1$&    \\ \cline{2-7}
\multicolumn{1}{ |c  }{}                        &
\multicolumn{1}{ |c| }{$P(nr_{100}>unp)$} & $1$ & $1$ & $1$ & $0.91$ & $0.99$ &   \\ \cline{1-7}
\multicolumn{1}{ |c  }{\multirow{2}{*}{Color} } &
\multicolumn{1}{ |c| }{$P(nr_{85}>unp)$} & $1$ & $0.99$ & $0.99$ & $0.99$ & $0.99$ \\ \cline{2-7}
\multicolumn{1}{ |c  }{}                        &
\multicolumn{1}{ |c| }{$P(nr_{100}>unp)$} & $0.98$ & $0.91$ & $0.89$ & $0.24$ & $0.27$ \\ \cline{1-7}
\multicolumn{1}{ |c  }{\multirow{2}{*}{Letter} } &
\multicolumn{1}{ |c| }{$P(nr_{85}>unp)$} & $1$ & $1$ & $1$ & $0.96$ & $0.99$ \\ \cline{2-7}
\multicolumn{1}{ |c  }{}                        &
\multicolumn{1}{ |c| }{$P(nr_{100}>unp)$} & $1$ & $0.44$ & $0.99$ & $0.22$ & $0.19$ \\ \cline{1-7}
\multicolumn{1}{ |c  }{\multirow{2}{*}{Mean} } &
\multicolumn{1}{ |c| }{$P(nr_{85}>unp)$} & $1$ & $1$ & $1$ & $1$ & $1$ \\ \cline{2-7}
\multicolumn{1}{ |c  }{}                        &
\multicolumn{1}{ |c| }{$P(nr_{100}>unp)$} & $1$ & $0.99$ & $1$ & $0.50$ & $0.87$ \\ \cline{1-7}
\end{tabular}
\label{table : Intelligibilty}
\end{table}

\section{Conclusion}
\label{sec:Disc and Conc}

In this paper, we have proposed a model-based method for performing binaural/bilateral speech enhancement in HAs. The proposed enhancement framework takes into account the speech production dynamics by using a FLKS for the enhancement process. The filter parameters required for the functioning of the FLKS are estimated jointly using the information at the left and right microphones. The filter parameters considered here are the speech and noise STP parameters and the speech pitch parameters. The estimation of these parameters in not trivial due to the highly non-stationary nature of speech and the noise in a cocktail party scenario. In this work, we have proposed a binaural codebook-based method, trained on spectral models of speech and noise, for estimating the speech and noise STP parameters, and a pitch estimator based on the harmonic model is proposed to estimate the pitch parameters. We then evaluated the proposed enhancement framework    in two experimental set-ups representative of the cocktail party scenario. The objective measures,  STOI and PESQ, were used for evaluating the proposed enhancement framework. The proposed method showed considerable improvement in STOI and PESQ scores, in comparison to a number of reference methods. Subjective listening tests when having access to single channel noisy observation also showed improvement in terms of intelligibility and quality. In the case of  intelligibility tests, a mean improvement of about $15$ \% was observed at -$8$ dB SNR.

\appendices
\section{Prediction and Correction stages of the FLKS}
\label{ssec:App_pred}
 This section gives the prediction and correction stages involved in the FLKS for the V-UV model. The same equations apply for the UV model, except that the state vector and the state transition matrices will be different. The prediction stage of the FLKS, which computes the a priori estimates of the state vector ($\hat{\bar{\mathbf{x}}}^{\text{\tiny{V-UV}}}_{l/r}(n|n-1)$) and error covariance matrix ($\mathbf{M}(n|n-1)$) is given by
\begin{equation*}
\label{eq:state prediction}
\hat{\bar{\mathbf{x}}}^{\text{\tiny{V-UV}}}_{l/r}(n|n-1) = \mathbf{F}^{\text{\tiny{V-UV}}}(f_n)\hat{\bar{\mathbf{x}}}^{\text{\tiny{V-UV}}}_{l/r}(n-1|n-1)
\end{equation*}
\begin{equation*}
\begin{split}
\label{eq:error_cov_prediction}
\mathbf{M}(n|n-1) = \mathbf{F}^{\text{\tiny{V-UV}}}(f_n)\mathbf{M}(n-1|n-1)\mathbf{F}^{\text{\tiny{V-UV}}}(f_n)^T + \\ \mathbf{\Gamma}_5 \begin{bmatrix}
\sigma_d^2(f_n) & 0 \\
0 & \sigma_v^2(f_n)
\end{bmatrix} \mathbf{\Gamma}_5^T.
\end{split}
\end{equation*}
The Kalman gain is computed as 
\begin{equation}
\label{eq: Kalman gain}
\mathbf{K}(n) = \frac{\mathbf{M}(n|n-1)\mathbf{\Gamma}^{{\text{\tiny{V-UV}}}}}{ [\mathbf{\Gamma}^{{\text{\tiny{V-UV}}}^T} \mathbf{M}(n|n-1) \mathbf{\Gamma}^{{\text{\tiny{V-UV}}}}]}.
\end{equation}
The correction stage of the FLKS, which computes the a posteriori estimates of the state vector and error covariance matrix is given by
\begin{equation*}
\label{eq: state correction}
\scalebox{0.9}{$
\hat{\bar{\mathbf{x}}}^{\text{\tiny{V-UV}}}_{l/r}(n|n) = \hat{\bar{\mathbf{x}}}^{\text{\tiny{V-UV}}}_{l/r}(n|n-1) + \mathbf{K}(n)[z_{l/r}(n) - \mathbf{\Gamma}^{{\text{\tiny{V-UV}}}^T}\hat{\bar{\mathbf{x}}}^{\text{\tiny{V-UV}}}_{l/r}(n|n-1)] $}
\end{equation*}
\begin{equation*}
\mathbf{M}(n|n) = (\mathbf{I} - \mathbf{K}(n) \mathbf{\Gamma}^{{\text{\tiny{V-UV}}}^T})\mathbf{M}(n|n-1).
\end{equation*}
Finally, the enhanced signal at time index $n-(d_s+1)$ is obtained by taking the $(d_s+1)^{th}$ entry of the a posteriori estimate of the state vector as 
\begin{equation}
\label{eq:final_enhanced_signal}
\hat{s}_{l/r}(n-(d_s+1)) = \left[ \hat{\bar{\mathbf{x}}}^{\text{\tiny{V-UV}}}_{l/r}(n|n) \right]_{d_s+1}.
\end{equation} 

\section{Behaviour of the likelihood function}
\label{ssec:behav}
\begin{figure}[!htbp]
\centering
\includegraphics[width=3.6cm, height=3.5cm]{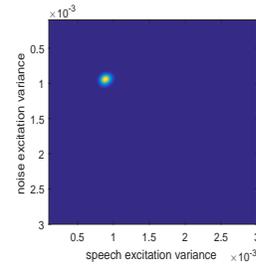}
\caption{Likelihood shown as a function of the speech and noise excitation variance.}
\label{fig: likelihood}
\end{figure}

For a given set of speech and noise AR coefficients, we show the behaviour of the likelihood $p(\mathbf{z}_l,\mathbf{z}_r|\boldsymbol{\theta})$ as a function of the speech and noise excitation variance. For the experiments, we have set the excitation variances to be $10^{-3}$. Fig. \ref{fig: likelihood} plots the likelihood as a function of the speech and noise excitation variance. It can be seen from the figure that likelihood is the maximum at the true values and decays rapidly as it deviates form its true value. This behaviour motivates the approximation in Section \ref{ssec : codebook}. 

\section{A priori information on the distribution of the  excitation variances}
\label{ssec: gamma}
\begin{figure}[h]
\centering
\includegraphics[width=5.0cm, height=3cm]{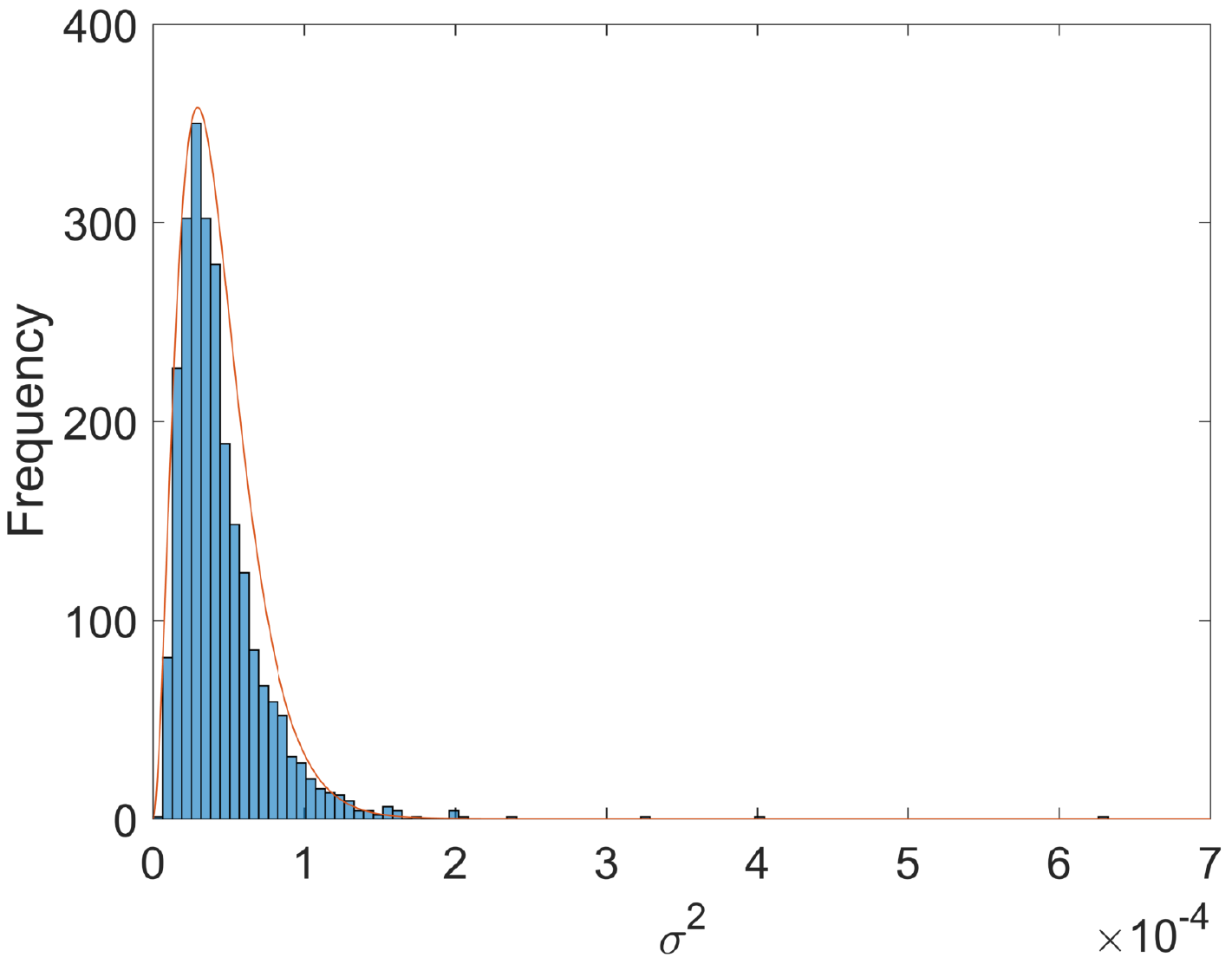}
\caption{Plot showing the histogram fitting for noise excitation variance. Curve (red) is obtained by fitting the histogram with a Gamma distribution with two parameters.}
\label{fig: histogram_exc_var}
\end{figure}

It can be seen from (\ref{eq: final_discrete_MMSE}) that the prior distributions of the excitation variances are used in the estimation of STP parameters. In the case of no a priori knowledge regarding the excitation variances, a uniform distribution can be used as done in \cite{srinivasan2007codebook}, but a priori knowledge regarding the distribution of  the noise excitation variance can be beneficial. Fig. \ref{fig: histogram_exc_var} shows the histogram of the noise excitation variance plotted for a minute of babble noise \cite{hirsch2000aurora}. It can be observed from the figure that the histogram approximately follows a Gamma distribution. Thus, we here use a Gamma distribution to model the a priori information about the noise excitation variance, which is 
 modelled using two parameters (shape parameter $\kappa$ and the scale parameter $\zeta$) as 
\begin{equation}
p(\sigma_v^2) = \frac{1}{\Gamma(\kappa)\zeta^k}\sigma_v^{2^{\kappa-1}}e^{-\frac{\sigma_v^2}{\zeta}},
\end{equation}
where  $\Gamma(\cdot)$ is the Gamma function. The parameters $\zeta$ and $\kappa$ can be learned from the training data.
 
\section*{Acknowledgment}

The authors would like to thank Innovation Fund Denmark (Grant No. 99-2014-1) for the financial support.

\ifCLASSOPTIONcaptionsoff
  \newpage
\fi




%

\bibliographystyle{IEEEtran}
\bibliography{BibTex/IEEEabrv,BibTeX/myabrv,mybib}

\begin{thebibliography}{10}
\providecommand{\url}[1]{#1}
\csname url@samestyle\endcsname
\providecommand{\newblock}{\relax}
\providecommand{\bibinfo}[2]{#2}
\providecommand{\BIBentrySTDinterwordspacing}{\spaceskip=0pt\relax}
\providecommand{\BIBentryALTinterwordstretchfactor}{4}
\providecommand{\BIBentryALTinterwordspacing}{\spaceskip=\fontdimen2\font plus
\BIBentryALTinterwordstretchfactor\fontdimen3\font minus
  \fontdimen4\font\relax}
\providecommand{\BIBforeignlanguage}[2]{{%
\expandafter\ifx\csname l@#1\endcsname\relax
\typeout{** WARNING: IEEEtran.bst: No hyphenation pattern has been}%
\typeout{** loaded for the language `#1'. Using the pattern for}%
\typeout{** the default language instead.}%
\else
\language=\csname l@#1\endcsname
\fi
#2}}
\providecommand{\BIBdecl}{\relax}
\BIBdecl

\bibitem{kochkin200210}
S.~Kochkin, ``10-year customer satisfaction trends in the {US} hearing
  instrument market,'' \emph{Hearing Review}, vol.~9, no.~10, pp. 14--25, 2002.

\bibitem{van2009speech}
T.~V.~D.~Bogaert, S.~Doclo, J.~Wouters, and M.~Moonen, ``Speech enhancement
  with multichannel {W}iener filter techniques in multimicrophone binaural
  hearing aids,'' \emph{The Journal of the Acoustical Society of America}, vol.
  125, no.~1, pp. 360--371, 2009.

\bibitem{bronkhorst1988effect}
A.~Bronkhorst and R.~Plomp, ``The effect of head-induced interaural time and
  level differences on speech intelligibility in noise,'' \emph{The Journal of
  the Acoustical Society of America}, vol.~83, no.~4, pp. 1508--1516, 1988.

\bibitem{doclo2008acoustic}
S.~Doclo, S.~Gannot, M.~Moonen, and A.~Spriet, ``Acoustic beamforming for
  hearing aid applications,'' \emph{Handbook on array processing and sensor
  networks}, pp. 269--302, 2008.

\bibitem{cornelis2010theoretical}
B.~Cornelis, S.~Doclo, T.~Van~dan Bogaert, M.~Moonen, and J.~Wouters,
  ``Theoretical analysis of binaural multimicrophone noise reduction
  techniques,'' \emph{{IEEE} Trans. Audio, Speech, and Language Process.},
  vol.~18, no.~2, pp. 342--355, 2010.

\bibitem{klasen2007binaural}
T.~J. Klasen, T.~V.~D.~Bogaert, M.~Moonen, and J.~Wouters, ``Binaural noise
  reduction algorithms for hearing aids that preserve interaural time delay
  cues,'' \emph{IEEE Trans. on Signal Process.}, vol.~55, no.~4, pp.
  1579--1585, 2007.

\bibitem{dorbecker1996combination}
M.~Dorbecker and S.~Ernst, ``Combination of two-channel spectral subtraction
  and adaptive {W}iener post-filtering for noise reduction and
  dereverberation,'' in \emph{Signal Processing Conference, 1996
  European}.\hskip 1em plus 0.5em minus 0.4em\relax IEEE, 1996, pp. 1--4.

\bibitem{li2011two}
J.~Li, S.~Sakamoto, S.~Hongo, M.~Akagi, and Y.~Suzuki, ``Two-stage binaural
  speech enhancement with {W}iener filter for high-quality speech
  communication,'' \emph{Speech Communication}, vol.~53, no.~5, pp. 677--689,
  2011.

\bibitem{lotter2006dual}
T.~Lotter and P.~Vary, ``Dual-channel speech enhancement by superdirective
  beamforming,'' \emph{EURASIP Journal on Advances in Signal Processing}, vol.
  2006, no.~1, pp. 1--14, 2006.

\bibitem{paliwal1987kalman}
K.~K. Paliwal and A.~Basu, ``A speech enhancement method based on {K}alman
  filtering,'' \emph{Proc. Int. Conf. Acoustics, Speech, Signal Processing},
  1987.

\bibitem{gibson1991filtering}
J.~D. Gibson, B.~Koo, and S.~D. Gray, ``Filtering of colored noise for speech
  enhancement and coding,'' \emph{IEEE Trans. Signal Process.}, vol.~39, no.~8,
  pp. 1732--1742, 1991.

\bibitem{gannot1998iterative}
S.~Gannot, D.~Burshtein, and E.~Weinstein, ``Iterative and sequential {K}alman
  filter-based speech enhancement algorithms,'' \emph{{IEEE} Trans. Acoust.,
  Speech, Signal Process.}, vol.~6, no.~4, pp. 373--385, 1998.

\bibitem{goh1999kalman}
Z.~Goh, K.~C. Tan, and B.~T.~G. Tan, ``Kalman-filtering speech enhancement
  method based on a voiced-unvoiced speech model,'' \emph{{IEEE} Trans.
  Acoust., Speech, Signal Process.}, vol.~7, no.~5, pp. 510--524, 1999.

\bibitem{srinivasan2007codebook}
S.~Srinivasan, J.~Samuelsson, and W.~B. Kleijn, ``Codebook-based {B}ayesian
  speech enhancement for nonstationary environments,'' \emph{{IEEE} Trans.
  Audio, Speech, and Language Process.}, vol.~15, no.~2, pp. 441--452, 2007.

\bibitem{mathew2016}
M.~S. Kavalekalam, M.~G. Christensen, F.~Gran, and J.~B. Boldt, ``Kalman filter
  for speech enhancement in cocktail party scenarios using a codebook based
  approach,'' \emph{Proc. Int. Conf. Acoustics, Speech, Signal Processing},
  2016.

\bibitem{mathew2016iwaenc}
M.~S. Kavalekalam, M.~G. Christensen, and J.~B. Boldt, ``Binaural speech
  enhancement using a codebook based approach,'' \emph{Proc. Int. Workshop on
  Acoustic Signal Enhancement}, 2016.

\bibitem{kavalekalam2017model}
------, ``Model based binaural enhancement of voiced and unvoiced speech,''
  \emph{Proc. Int. Conf. Acoustics, Speech, Signal Processing}, 2017.

\bibitem{makhoul1975linear}
J.~Makhoul, ``Linear prediction: A tutorial review,'' \emph{Proceedings of the
  IEEE}, vol.~63, no.~4, pp. 561--580, 1975.

\bibitem{he2017multiplicative}
Q.~He, F.~Bao, and C.~Bao, ``Multiplicative update of auto-regressive gains for
  codebook-based speech enhancement,'' \emph{{IEEE} Trans. Audio, Speech, and
  Language Process.}, vol.~25, no.~3, pp. 457--468, 2017.

\bibitem{christopher2016pattern}
M.~B. Christopher, \emph{Pattern recognition and machine learning}.\hskip 1em
  plus 0.5em minus 0.4em\relax Springer-Verlag New York, 2006.

\bibitem{gray2006toeplitz}
R.~M. Gray \emph{et~al.}, ``Toeplitz and circulant matrices: A review,''
  \emph{Foundations and Trends{\textregistered} in Communications and
  Information Theory}, vol.~2, no.~3, pp. 155--239, 2006.

\bibitem{itakura1968analysis}
F.~Itakura, ``Analysis synthesis telephony based on the maximum likelihood
  method,'' in \emph{The 6th international congress on acoustics, 1968}, 1968,
  pp. 280--292.

\bibitem{lee2001algorithms}
D.~D. Lee and H.~S. Seung, ``Algorithms for non-negative matrix
  factorization,'' in \emph{Advances in neural information processing systems},
  2001, pp. 556--562.

\bibitem{fevotte2009nonnegative}
C.~F{\'e}votte, N.~Bertin, and J.-L. Durrieu, ``Nonnegative matrix
  factorization with the {I}takura-{S}aito divergence: With application to
  music analysis,'' \emph{Neural computation}, vol.~21, no.~3, pp. 793--830,
  2009.

\bibitem{stoica2005spectral}
P.~Stoica, R.~L. Moses \emph{et~al.}, \emph{Spectral analysis of
  signals}.\hskip 1em plus 0.5em minus 0.4em\relax Pearson Prentice Hall Upper
  Saddle River, NJ, 2005, vol. 452.

\bibitem{kamkar2009improved}
A.~H. Kamkar-Parsi and M.~Bouchard, ``Improved noise power spectrum density
  estimation for binaural hearing aids operating in a diffuse noise field
  environment,'' \emph{{IEEE} Trans. Audio, Speech, and Language Process.},
  vol.~17, no.~4, pp. 521--533, 2009.

\bibitem{jeub2011robust}
M.~Jeub, C.~Nelke, H.~Kruger, C.~Beaugeant, and P.~Vary, ``Robust dual-channel
  noise power spectral density estimation,'' in \emph{Signal Processing
  Conference, 2011 19th European}.\hskip 1em plus 0.5em minus 0.4em\relax IEEE,
  2011, pp. 2304--2308.

\bibitem{brown1998structural}
P.~C. Brown and R.~O. Duda, ``A structural model for binaural sound
  synthesis,'' \emph{{IEEE} Trans. Acoust., Speech, Signal Process.}, vol.~6,
  no.~5, pp. 476--488, 1998.

\bibitem{christensen2009multi}
M.~G. Christensen and A.~Jakobsson, ``Multi-pitch estimation,'' \emph{Synthesis
  Lectures on Speech \& Audio Processing}, vol.~5, no.~1, pp. 1--160, 2009.

\bibitem{cooke2006audio}
M.~Cooke, J.~Barker, S.~Cunningham, and X.~Shao, ``An audio-visual corpus for
  speech perception and automatic speech recognition,'' \emph{The Journal of
  the Acoustical Society of America}, vol. 120, no.~5, pp. 2421--2424, 2006.

\bibitem{linde1980algorithm}
Y.~Linde, A.~Buzo, and R.~M. Gray, ``An algorithm for vector quantizer
  design,'' \emph{IEEE Trans. Communications}, vol.~28, no.~1, pp. 84--95,
  1980.

\bibitem{gray1976distance}
A.~Gray and J.~Markel, ``Distance measures for speech processing,''
  \emph{{IEEE} Trans. Acoust., Speech, Signal Process.}, vol.~24, no.~5, pp.
  380--391, 1976.

\bibitem{etsi_binaural}
ETSI202396-1, ``Speech and multimedia transmission quality; part 1: Background
  noise simulation technique and background noise database.'' 2009.

\bibitem{kayser2009database}
H.~Kayser, S.~D. Ewert, J.~Anem{\"u}ller, T.~Rohdenburg, V.~Hohmann, and
  B.~Kollmeier, ``Database of multichannel in-ear and behind-the-ear
  head-related and binaural room impulse responses,'' \emph{EURASIP Journal on
  Advances in Signal Processing}, vol. 2009, no.~1, pp. 1--10, 2009.

\bibitem{wabnitz2010room}
A.~Wabnitz, N.~Epain, C.~Jin, and A.~Van~Schaik, ``Room acoustics simulation
  for multichannel microphone arrays,'' in \emph{Proceedings of the
  International Symposium on Room Acoustics}, 2010, pp. 1--6.

\bibitem{erkelens2007minimum}
J.~S. Erkelens, R.~C. Hendriks, R.~Heusdens, and J.~Jensen, ``Minimum
  mean-square error estimation of discrete fourier coefficients with
  generalized gamma priors,'' \emph{{IEEE} Trans. Audio, Speech, and Language
  Process.}, vol.~15, no.~6, pp. 1741--1752, 2007.

\bibitem{loizou2005speech}
P.~C. Loizou, ``Speech enhancement based on perceptually motivated {B}ayesian
  estimators of the magnitude spectrum,'' \emph{{IEEE} Trans. Acoust., Speech,
  Signal Process.}, vol.~13, no.~5, pp. 857--869, 2005.

\bibitem{taal2011algorithm}
C.~H. Taal, R.~C. Hendriks, R.~Heusdens, and J.~Jensen, ``An algorithm for
  intelligibility prediction of time--frequency weighted noisy speech,''
  \emph{{IEEE} Trans. Audio, Speech, and Language Process.}, vol.~19, no.~7,
  pp. 2125--2136, 2011.

\bibitem{recommendation2001perceptual}
``Perceptual evaluation of speech quality, an objective method for end-to-end
  speech quality assessment of narrowband telephone networks and speech
  codecs,'' \emph{ITU-T Recommendation}, p. 862, 2001.

\bibitem{falk2015objective}
T.~H. Falk, V.~Parsa, J.~F. Santos, K.~Arehart, O.~Hazrati, R.~Huber, J.~M.
  Kates, and S.~Scollie, ``Objective quality and intelligibility prediction for
  users of assistive listening devices: Advantages and limitations of existing
  tools,'' \emph{IEEE signal processing magazine}, vol.~32, no.~2, pp.
  114--124, 2015.

\bibitem{bronkhorst1990clinical}
A.~W. Bronkhorst and R.~Plomp, ``A clinical test for the assessment of binaural
  speech perception in noise,'' \emph{Audiology}, vol.~29, no.~5, pp. 275--285,
  1990.

\bibitem{recommendation20031534}
I.~Recommendation, ``1534-1: Method for the subjective assessment of
  intermediate quality level of coding systems,'' \emph{International
  Telecommunication Union}, 2003.

\bibitem{hirsch2000aurora}
H.-G. Hirsch and D.~Pearce, ``The aurora experimental framework for the
  performance evaluation of speech recognition systems under noisy
  conditions,'' in \emph{ASR2000-Automatic Speech Recognition: Challenges for
  the new Millenium ISCA Tutorial and Research Workshop (ITRW)}, 2000.

\bibitem{anzalone2006determination}
M.~C. Anzalone, L.~Calandruccio, K.~A. Doherty, and L.~H. Carney,
  ``Determination of the potential benefit of time-frequency gain
  manipulation,'' \emph{Ear and hearing}, vol.~27, no.~5, p. 480, 2006.

\bibitem{loizou2011reasons}
P.~C. Loizou and G.~Kim, ``Reasons why current speech-enhancement algorithms do
  not improve speech intelligibility and suggested solutions,'' \emph{{IEEE}
  Trans. Audio, Speech, and Language Process.}, vol.~19, no.~1, pp. 47--56,
  2011.

\bibitem{kolbaek2017speech}
M.~Kolb{\ae}k, Z.-H. Tan, and J.~Jensen, ``Speech intelligibility potential of
  general and specialized deep neural network based speech enhancement
  systems,'' \emph{{IEEE} Trans. Audio, Speech, and Language Process.},
  vol.~25, no.~1, pp. 153--167, 2017.

\bibitem{nielsen2017fast}
J.~K. Nielsen, T.~L. Jensen, J.~R. Jensen, M.~G. Christensen, and S.~H. Jensen,
  ``Fast fundamental frequency estimation: Making a statistically efficient
  estimator computationally efficient,'' \emph{Signal Processing}, vol. 135,
  pp. 188--197, 2017.

\end{thebibliography}

\begin{IEEEbiography}[{\includegraphics[width=1in,height=2in,clip,keepaspectratio]{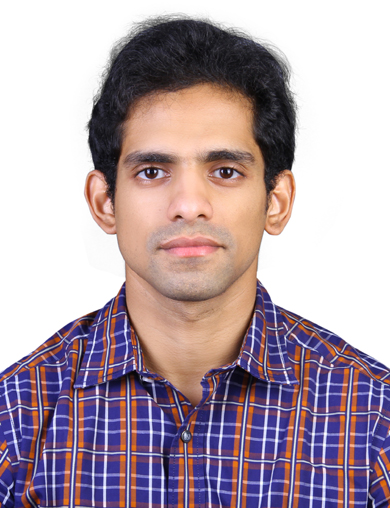}}]{Mathew Shaji Kavalekalam}
was born in Thrissur, India in 1989. He received his B.Tech in electronics and communications engineering from Amrita University and M.Sc in communications engineering from RWTH Aachen university in 2011 and 2014 respectively. He is currently a PhD student at the Audio Analysis Lab, Department of Architecture, Design and Media Technology, Aalborg University. His research interests include speech enhancement for Hearing aid applications.
\end{IEEEbiography}
\begin{IEEEbiography}[{\includegraphics[width=1in,height=1.25in,clip,keepaspectratio]{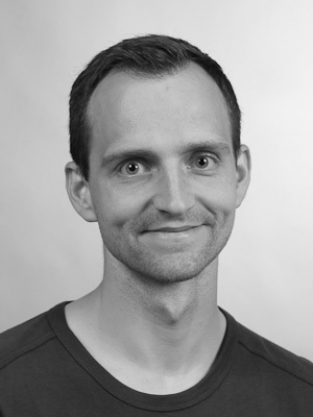}}]{Jesper Kj\ae r Nielsen} (S'12--M'13) received the M.Sc (Cum Laude) and Ph.D. degrees in electrical engineering with a specialisation in signal processing from Aalborg University, Denmark, in 2009 and 2012, respectively. From 2012 to 2016, he was with the Department of Electronic Systems, Aalborg University, as an industrial postdoctoral researcher (12-15) and as a non-tenured associate professor (15-16). Bang \& Olufsen A/S (B\&O) was the industrial partner in these four years. Jesper is currently with the Audio Analysis Lab, Aalborg University, in a three year position as an assistant professor in Statistical Signal Processing. He is part-time employed by B\&O and part time employed on a research project with the Danish hearing aid company GN ReSound.

Jesper has been a Visiting Scholar in the Signal Processing and Communications Laboratory, University of Cambridge in 2009 and at the Department of Computer Science, University of Illinois at Urbana-Champaign in 2011. Moreover, he has been a guest researcher in the Signal \& Information Processing Lab at TU Delft in 2014. His research interests include spectral estimation, (sinusoidal) parameter estimation, microphone array processing, as well as statistical and Bayesian methods for signal processing.
\end{IEEEbiography}

\begin{IEEEbiography}[{\includegraphics[width=1in,height=2in,clip,keepaspectratio]{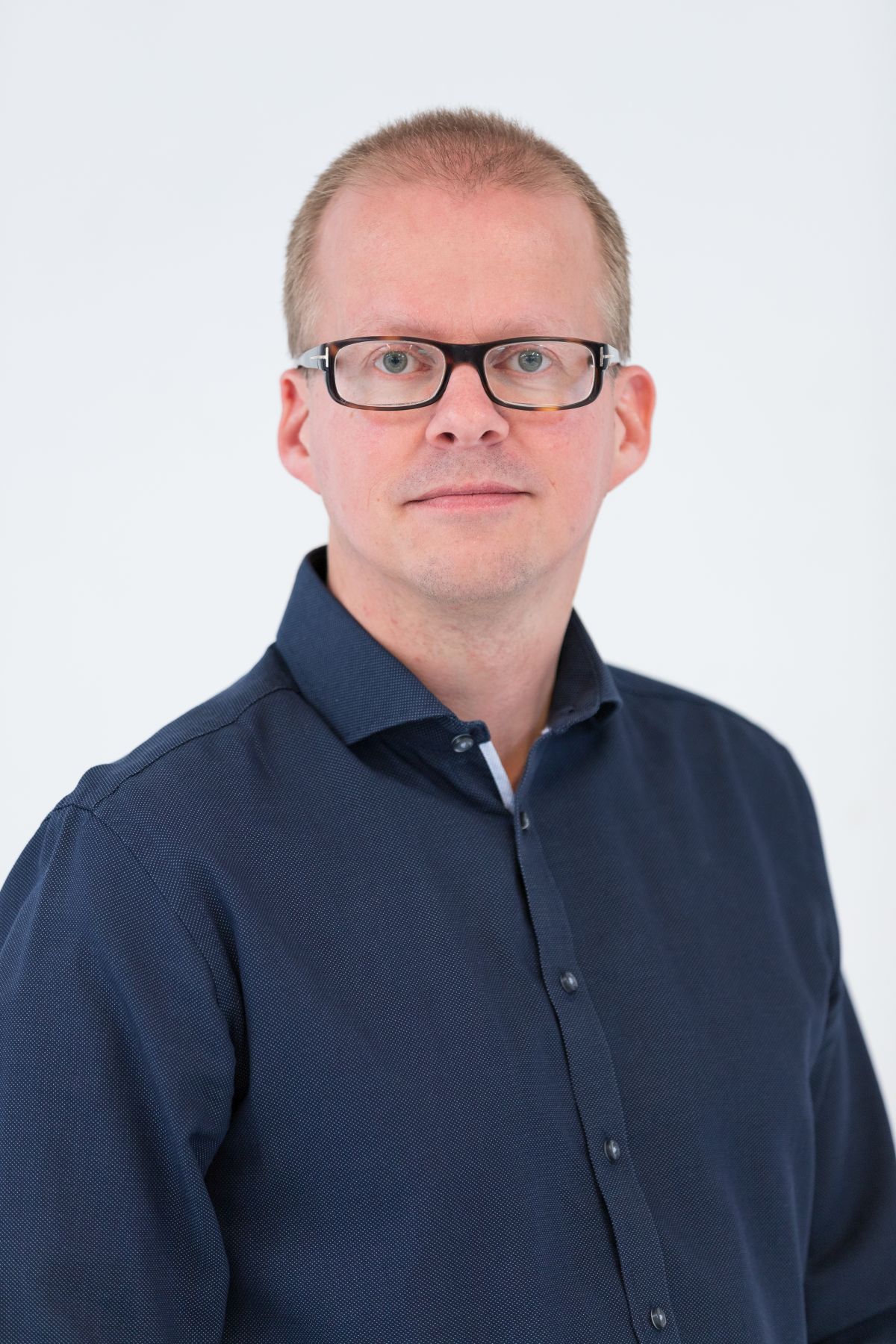}}]{Jesper B\"unsow Boldt}
 received the M.Sc. degree in Electrical Engineering in 2003 and the Ph.D. degree in Signal Processing in 2010, both from Aalborg University (AAU) in Denmark. After his Master’s studies he joined Oticon as Hearing Aid Algorithm Developer and from 2007 as Industrial Ph.D. Researcher jointly with Aalborg University and the Technical University of Denmark (DTU). He has been visiting researcher at both Columbia University and Eriksholm Research Centre. In 2013 he joined GN ReSound as Senior Research Scientist and in 2015 he became Research Team Manager in GN Advanced Science.

His main interest is the cocktail party problem and the research that has the potential to solve this problem for hearing impaired individuals. This includes speech, audio, and acoustic signal processing but also auditory signal processing, psychoacoustics, and perception. 

\end{IEEEbiography}
\begin{IEEEbiography}[{\includegraphics[width=1in,height=2in,clip,keepaspectratio]{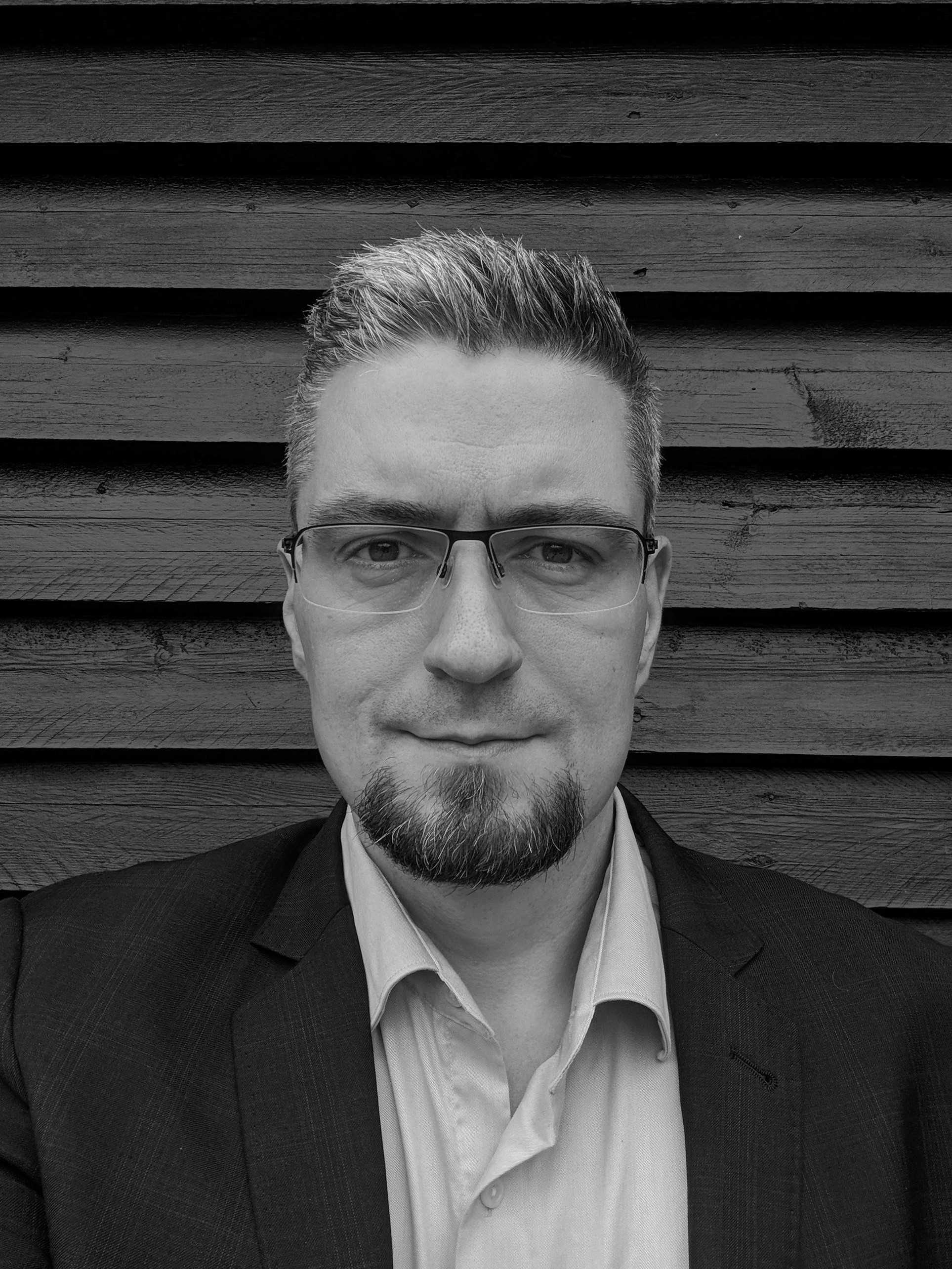}}]{Mads Gr{\ae}sb{\o}ll Christensen} (S'00--M'05--SM'11))
received the M.Sc. and Ph.D. degrees in 2002 and 2005, respectively, from Aalborg
University (AAU) in Denmark, where he is also currently employed at the
Dept. of Architecture, Design \& Media Technology as Professor in Audio Processing and is head and founder of the Audio Analysis Lab.

He was formerly with the Dept. of Electronic Systems at AAU and has been held visiting positions at Philips Research
Labs, ENST, UCSB, and Columbia University. He has published 3 books and more than 200 papers in peer-reviewed conference proceedings and journals, and he has given multiple tutorials at EUSIPCO, SMC, and INTERSPEECH and a keynote talk at IWAENC. His research interests lie in audio and acoustic signal processing where he has worked on topics such as microphone arrays, noise reduction, signal modeling, speech analysis, audio classification, and audio coding. 

Dr. Christensen has received several awards, including best paper awards, the Spar Nord Foundation’s Research Prize, a Danish Independent Research Council Young Researcher’s Award, the Statoil Prize, the EURASIP Early Career Award, and an IEEE SPS best paper award. He is a beneficiary of major grants from the Independent Research Fund Denmark, the Villum Foundation, and Innovation Fund Denmark. He is a former Associate Editor for IEEE/ACM Trans. on Audio, Speech, and Language Processing and IEEE Signal Processing Letters, a member of the IEEE Audio and Acoustic Signal Processing Technical Committee, and a founding member of the EURASIP Special Area Team in Acoustic, Sound and Music Signal Processing. He is Senior Member of the IEEE, Member of EURASIP, and Member of the Danish Academy of Technical Sciences.
\end{IEEEbiography}

\end{document}